\documentclass[aps,floats,twocolumn,epsf,prb,showpacs]{revtex4}
\usepackage{latexsym,amsmath,amssymb,bm,graphicx,amsfonts,bbold}
\usepackage{graphicx,color}
\usepackage{psfrag}
\usepackage{braket}
\usepackage{epsfig}
\makeatletter\def\Dated@name{}\makeatother

\newcommand{\deps}{\Delta\varepsilon}

\newcommand{\nna}{n$_\text{Na}$}
\newcommand{\nvac}{n$_\text{Vac}$}

\newcommand{\xna}{x_\text{Na}}
\newcommand{\xvac}{x_\text{Vac}}
\newcommand{\Cobaltate}{Na$_\text{x}$CoO$_2$}
\newcommand{\cobaltate}{N{a}$_{0.7}$CoO$_2$}
\newcommand{\Weiss}{\GG_0}
\newcommand{\kboltz}{k_\text{B}}
\newcommand{\GG}{{\cal G}}
\newcommand{\AAA}{{\mathcal A}}
\newcommand{\aquivalent}{\stackrel{\scriptscriptstyle\wedge}{=}}
\newcommand{\KK}{{\cal K}}
\newcommand{\eLDA}{\varepsilon^{\text{LDA}}}
\newcommand{\eTB}{\varepsilon^{\text{TB}}}
\newcommand{\vv}{{\vec{v}}}
\newcommand{\CoNa}{Co$_\text{Na}$}
\newcommand{\CoVac}{Co$_\text{Vac}$}
 \providecommand{\abs}[1]{\lvert#1\rvert}
\newcommand{\OO}{{\cal O}}
\newcommand{\PP}{{\cal P}}

\newcommand{\hj}{\widehat j}
\newcommand{\hO}{\widehat O}
\newcommand{\dpsi}{\dot\psi}

\newcommand{\gv}{\nabla\varepsilon}
\begin{document}

\title{Effects of electronic correlations and disorder on the thermopower of Na$_x$CoO$_2$}
\author{P. Wissgott$^{1}$, A. Toschi$^{1}$, G. Sangiovanni$^1$ and K. Held$^1$}
\affiliation{$^1$ Institute for Solid State Physics, Vienna University of Technology,
1040 Vienna, Austria\\
%$^2$ University of Electro-Communications 1-5-1 Chofugaoka, Chofu-shi
%Tokyo 182-8585, Japan
}
\date{\today}

\begin{abstract}
For the thermoelectric properties of Na$_x$CoO$_2$,
we analyze  the effect of local Coulomb interaction and (disordered) potential
differences
for Co-sites with adjacent Na-ion or vacancy.   The disorder potential alone
increases the resistivity and reduces the thermopower, while the Coulomb
interaction
alone leads only to minor changes compared to the one-particle picture
of the local density approximation. Only combined, these two
terms give rise to a substantial increase of the thermopower:
the number of (quasi-)electrons around the Fermi level
is much more suppressed than that of the (quasi-)holes.
Hence, there is a particle-hole imbalance acting in the same direction
as a similar imbalance for the group velocities. Together, this interplay
results in a large positive thermopower.
Introducing a thermoelectric spectral density, we located
the energies and momenta regions most relevant for the
thermopower and changes thereof.
%Combining band structure and correlation engineering might hence
%pave the route to substantially  increase the thermoelectric figure of merit.
\end{abstract}

\maketitle
\begin{figure*}[tb]
  \centering
\includegraphics[height=4.75 cm]{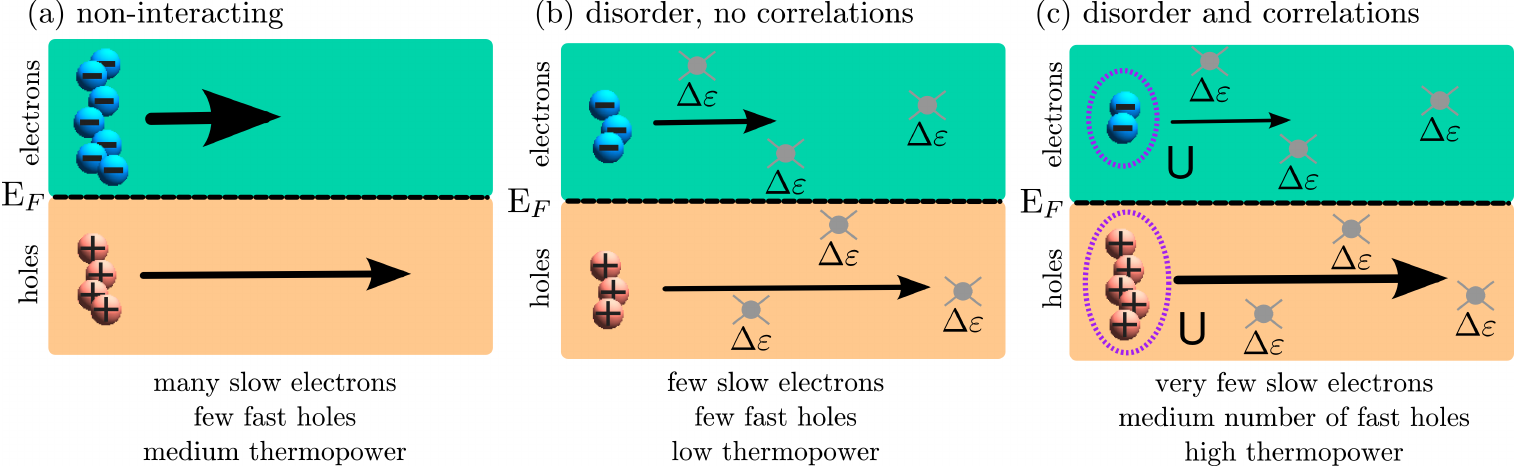}

   \caption{(Color online) Qualitative effect of disorder and electronic correlations on the charge carriers~\cite{quasi} and the thermopower in \Cobaltate. On the vertical axis, energy in arbitrary units is shown with the Fermi level E$_{F}$ and electrons above and holes below. The horizontal axis schematically indicates the movement through the material with velocities and charge carrier density according to the length and width of the arrow, respectively. The presence of disorder and electronic correlations is visualized by scattering centers $\deps$ and the encirclement of the charge carriers accompanied by $U$, respectively.}
    \label{Fig:Summary1}
\end{figure*}

\begin{figure}[tp]
  \centering
      \includegraphics[height=5cm]{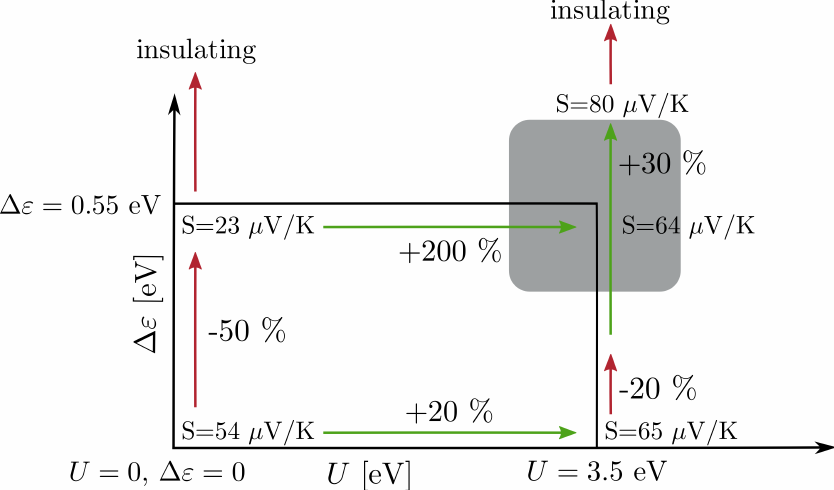}
\caption{(Color online) Change of the thermopower $S$ in the parameter space of local Coulomb repulsion $U$ and disorder potential $\deps$ for \cobaltate\; at $T=290$ K. The highlighted regions indicates realistic parameter values for this material.}
    \label{Fig:Parameter1}
\end{figure}
The efficient conversion between different forms of energy represents a primary task in technical applications. Thereby the arguably most valuable form is electrical energy. On the other hand, heat energy is widely used in power plants to produce mechanical energy which is then further converted to electricity. The reason for the widespread application of this multiple-step process is based on the lack of an efficient direct alternative.\\
A direct heat-electricity conversion is possible by the thermoelectric effect or thermopower. The~(dimensionless) thermoelectric figure of merit $ZT$ has been established as a benchmark to compare different materials. Values of $ZT\sim3$ appear to be necessary to trigger an extensive industrial use of thermoelectrics\cite{MRS,solar} which could then not only replace traditional methods of energy production but could also increase the efficiency of existing technical implementations such as cars\cite{cars} or power plants via the conversion of excess heat to electricity.\\
The development of novel thermoelectric applications follows two routes: First, there is the search for new materials with a large intrinsic $ZT$ which can either be achieved by optimizing features of already known compounds ~(via e.g. bandstructure engineering~\cite{PbTeDOS}) or by the discovery of new classes of materials with peculiar properties~(e.g. Kondo systems~\cite{Paschen,TEPAM1}). Second, one tries to diminish phonon losses of the thermopower which can be done by avoiding collective excitations of the crystal by a specific structural arrangement~(phonon engineering) such as heterostructuring~\cite{hetero1,hetero2}.\\
Opposed to other transport properties, as for example the electrical conductivity, the magnitude of the thermoelectric effect depends on an asymmetry between the two charge carriers holes and electrons. This asymmetry usually manifests itself in different velocities and/or spectral densities in the vicinity of the chemical potential.\\
In this paper, we investigate the thermoelectric properties of Na$_x$CoO$_2$ taken as a representative compound where the already favorable single-electron features are enhanced by the combination of strong electronic correlations and disorder. We extend the analysis of a previous work~\cite{short} exploring the dependence of the thermopower on various
parameters, one at a time as well as combinations of two of them,
e.g. disorder and correlation strength at the same time. \\
As a non-interacting starting point we used the local density approximation (LDA) result from Ref.~\onlinecite{Singh00} fitted to a tight-binding bandstructure\cite{short}. Already at the level of LDA the thermoelectric performance
of Na$_x$CoO$_2$ was found to be high and was understood in terms of the “pudding mold”
shape of the decisive $a_{1g}$ band~\cite{Kuroki07}. Here, we want to analyze the effects of electronic correlation in this compound and
therefore we have combined LDA with dynamical mean field theory (DMFT)~\cite{DMFT1,DMFT2}. The LDA+DMFT approach is described
in Refs.~\onlinecite{DMFT3,LDADMFT0,LDADMFT1,LDADMFT4} and we proceed as in Ref.~\onlinecite{Held09} to compute the thermoelectric response. We focused on the low-energy excitations
which give the main contributions to the thermopower. Earlier LDA+DMFT studies
on the spectral properties of Na$_x$CoO$_2$ can be found in Refs.~\onlinecite{Ishida05,Lechermann05,Kotliar07,Liebsch08,Lechermann09}.\\
In Fig.~\ref{Fig:Summary1}, we show a summary of the main results described in this study: Considering neither disorder nor correlations, the charge carrier properties are described by the ``pudding mold'' band which favors electrons in terms of spectral density, but favors holes with respect to group velocity. The latter effect is larger so that hole transport~\cite{quasi} dominates and we get a positive value of the thermopower $S$, Fig.\ref{Fig:Summary1}~(left panel). Introducing disorder to the system increases the scattering rate of both charge carriers thus diminishing $S$, Fig.\ref{Fig:Summary1}~(middle panel). If we additionally take into account electronic correlations, the hole spectral density is recovered and can be even over-compensated whereas the number of electrons is further decreased which overall yields a high thermoelectric response, Fig.\ref{Fig:Summary1}~(right panel). For a more quantitative picture, refer to Fig.~\ref{Fig:Parameter1}, where the change of $S$ in the parameter space of correlation $U$ and disorder $\deps$ is visualized. It can be seen that drastic changes happen for example for $\deps=0.55$ eV where an increasing $U$ leads to an enhancement of $S$ by $200\%$. Moreover, for even larger values of $\deps$ and $U=3.5$ eV we found the maximal values of $S~80$ $\mu$V/K in the tested parameter space. Note that the system shows insulating tendency for similar values of $\deps\gtrsim 0.55$ eV and $U=0$. This implies that in correlated materials disorder effects may be exploited to enhance the thermopower.\\
The structure of this paper is as follows: In Sec.~\ref{Sec:LDA+DMFT}, we introduce the Hamiltonian of the model and describe how standard LDA+DMFT can be expanded by means of coherent potential approximation~(CPA) to include disorder. In Sec.~\ref{Sec:LinearResponse}, we depict the linear response formalism to compute the thermopower $S$ as a functional of spectral density, group velocity and temperature. 
Both sections present the necessary details for following our result which were naturally missing in the short paper Ref.~\onlinecite{short}

Sec.~\ref{Sec:StructureProp} provides an insight to the structural and spectral properties of \Cobaltate\; missing in Ref.~\onlinecite{short}. Additionally to the lattice structure, the employed tight-binding approximation is explained in detail. Confirming previous results from Ref.~\onlinecite{Kuroki07}, the third nearest neighbor hopping appears to play a key role for the transport properties of this triangular compound. Disentangling the spectral contributions and the self energy from the two non-equivalent lattice sites due to disorder, it is shown that one of the two sites exhibits a much larger correlation effect than the other. A comparison of low energy spectral properties to experiment concludes this section. In Sec.\ref{Sec:Thermopower}, the main results for the thermopower are presented. The method of visualization of spectral contributions to the thermopower introduced in Ref.~\onlinecite{short}, is described in detail as well as new ways to simplify the theoretical thermoelectric analysis. In an effort to distinguish between the parameters temperature, disorder and electronic correlation, we investigate their influence on the thermopower with the respective other parameters fixed. It is revealed that the system is very sensitive to small changes of disorder. The impact of  correlations on the transport properties is directly connected to the strength of the disorder which can be seen in a detailed spectral analysis. Towards the end of this section, the dependence of the thermopower on a change of the sodium doping is delineated. Increasing the doping within the investigated doping region appears to increase the thermoelectric effect which is also supported by experimental results. Finally, we give a conclusion in Sec.~\ref{Sec:Conclusion}.

% LDA+DMFT Introduction, reference(Georges)
% In the parameter-space of [$U$,$\deps$] the thermopower $S$ varies strongly for fixed disorder $\deps=0.55$ eV~(realistic value,\cite{Kotliar07})

% \newpage
\section{LDA+DMFT approach including CPA}\label{Sec:LDA+DMFT}
\begin{figure}[tp]
  \centering
\includegraphics[height=6.5 cm]{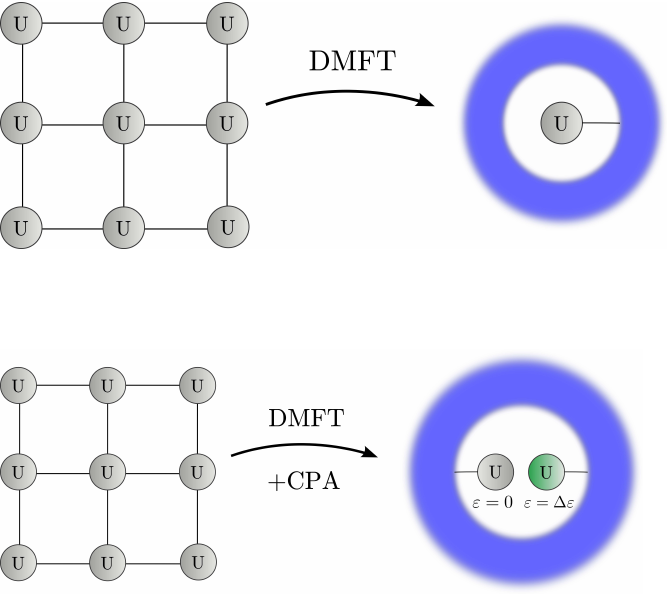}
\caption{(Color online) Schematic description of the dynamical mean field method~(DMFT) compared to its combination with the coherent potential approximation~(CPA). Top: In standard DMFT the lattice problem is replaced by a site in an self-consistently determined dynamical mean field. Bottom: For the CPA, \emph{two} sites with different on-site potential are connected to the same dynamical mean field.}
    \label{Fig:DMFTCPA1}
\end{figure}

We aim to model \Cobaltate\; with binary disorder, which corresponds to consider two different sites $a$,$b$ with an occurrence of $x_a,x_b=1-x_a$.
In comparison to the standard one-band Hubbard model~\cite{Hubbard63}, the disorder Hamiltonian
\begin{align}\label{Eq:DisorderHamiltonian}
 H_\text{dis}=-\sum_{ ij,\sigma}t_{ij}c^{+}_{j\sigma} c_{i\sigma}^{\phantom{+}} + U\sum_{ i} n_{i\uparrow}n_{i\downarrow}+\deps\sum_{i\in b} n_i
\end{align}
includes a term contributing only on the site $b$ by a disorder potential
$\deps$. Here, $c_{i\sigma}^+$ and $c_{i\sigma}$ create and annihilate an electron on site $i$ with spin $\sigma$, respectively, $n_{i\sigma}:=c_{i\sigma}^+c_{i\sigma}$ is the electron density on site $i$ for the spin $\sigma$, $U$ is the local Coulomb interaction. \\
For disordered systems the Coherent Potential Approximation~(CPA)~\cite{Elliot73,Vlaming91} provides a way to treat inequivalent sites in the same spirit as DMFT. In the CPA, the sites $a,b$ are assumed to be surrounded by a self-consistently determined effective medium. Both sites experience the same dynamical mean field, i.e. see the same~(averaged) environment given by an impurity Greens function~$\GG_0^{-1}$ as defined in Ref.~\onlinecite{DMFT2} or equivalently the same~(coherent) potential represented by a self energy $\Sigma$. Even though the mean field $\GG_0^{-1}$ is the same, the disorder manifests in two different on-site energies $\varepsilon_a=0,\varepsilon_b=\deps$ on the sites $a$ and $b$, respectively~(Fig.~\ref{Fig:DMFTCPA1}). \\
In DMFT the dynamical mean field $\GG_0^{-1}$ determines an action
\begin{equation}\label{Eq:DMFT1}
\begin{aligned}
% G(\tau)=&-\frac{1}{z}\int \DD\psi \DD\psi^+ \psi(\tau)\ \psi(0)^+ e^{\AAA[\psi,\psi^+,\GG_0^{-1}]},\\
%Z:=&\int \DD\psi \DD\psi^+ e^{\AAA[\psi,\psi^+,\GG_0^{-1}]},\\
\AAA:=& \sum_{\sigma\omega_m}\psi_\sigma^+(\omega_m) \GG_0^{-1}(\omega_m)\psi_\sigma(\omega_m)\\ &-U\sum_{\sigma}\int_{0}^\beta d\tau'\psi_\sigma^+(\tau')\psi_\sigma(\tau')\psi_{\sigma'}^+(\tau')\psi_{\sigma'}(\tau').
\end{aligned}
\end{equation}
where $\psi^+$, $\psi$ are creation and annihilation operators, respectively, $\sigma$ is a spin index, $\beta$ is the inverse temperature and $\omega_m$ are Matsubara frequencies~(see e.g. Ref.~\onlinecite{LDADMFT4}). The inclusion of (disorder)~CPA into DMFT is straightforward~(see Fig.~\ref{Fig:DMFTCPA1} for a schematic visualization): Assume for the moment that the effective medium $\GG_0^{-1}(i\omega_m)$ is known. Then, the local Greens functions $G_a(r=0,i\omega_m)$, $G_b(r=0,i\omega_m)$ can be determined via the action $\AAA$ from~(\ref{Eq:DMFT1}) by replacing
the standard $\GG_0^{-1}(\omega_m)$ by $\left(\GG_0^{-1}(\omega_m)-\mathbb{1}\varepsilon_{a,b}\right)$ with different on-site energies $\varepsilon_a=0,\varepsilon_b=\deps$. The local Greens functions $G_a,G_b$ are the local interacting propagators of electrons starting at $a,b$ and returning to the same site $a$ or $b$, respectively. Then, the total Greens function $G(r=0,i\omega_m)$ is given by the weighted mean
\begin{align}\label{Eq:Disorder1}
 G(i\omega_m)=x_a  G_{a}(i\omega_m)+(1-x_a)G_{b}(i\omega_m)
\end{align}
such that both propagators $G_a,G_b$ are taken into account according to the stoichiometric appearance of the corresponding site $x_a$ or $x_b$ in the lattice.\\
For the CPA expansion of the DMFT, the effective dynamic medium $\GG_0^{-1}(i\omega_m)$ is determined self-consistently. For a given $\GG_0^{-1}$, we obtain the local Greens functions $G_a,G_b$ by two Hirsch-Fye quantum Monte Carlo~(HF-QMC) calculations with different on-site potentials $\epsilon_a=0,\epsilon_b=\deps$. Then, the weighted $G(i\omega_m)$ from Eq.~\eqref{Eq:Disorder1} can be used to compute the new self energy $\Sigma(\omega)$ or equivalently the new effective field $\GG_0^{-1}$ via Dyson's equation
\begin{align}\label{Eq:AIMG0}
\Sigma(i\omega_m) =  \GG^{-1}_{0}(i\omega_m)-G^{-1}(i\omega_m).
\end{align}
With this approach we can include both correlations and disorder in a dynamic mean field scheme.

\section{Linear response and DMFT expressions for the thermopower}\label{Sec:LinearResponse}
In linear response, the thermopower is defined as a ratio of two correlation functions $K_1$ and $K_0$
\begin{align}\label{Eq:S11}
           S =\frac{k_\textrm{B}}{T}\frac{K_1}{K_0}.
\end{align}
Here and throughout, $T$ denotes the temperature. The determination of the correlation functions usually requires a representation of current operators in terms of the electric field operators $\psi^+,\psi$. The electrical current in imaginary time reads
\begin{align}
                     %=&-\frac{1}{2m}\lim_{\Dtau\rightarrow 0}\frac{1}{\Dtau}\lim_{\tir\rightarrow r}\nabla_{\alpha,r}\big[ \psi^+(\tir,\tau+\Dtau)\psi(r,\tau) -\psi^+(\tir,\tau)\psi(r,\tau) \label{Eq:CurrentRepsresentation2}\\ \nonumber  &\hspace*{3 cm}+\psi^+(r,\tau)\psi(\tir,\tau+\Dtau) -\psi^+(r,\tau)\psi(\tir,\tau)\big],\\
   \hj(r,\tau)&=\frac{i\abs{e}}{2m}\left[\psi^+(r,\tau)\nabla\psi(r,\tau)- \nabla\psi^+(r,\tau)\psi(r,\tau)\right]\label{Eq:CurrentRepsresentation3},
   %&=\frac{i\abs{e}}{2m}\lim_{\tir\rightarrow r}(\nabla_{\beta,r}-\nabla_{\beta,\tir})\psi^+(\tir,\tau)\psi(r,\tau),\label{Eq:CurrentRepsresentation4}
\end{align}
which can be found in many textbooks~\cite{Mahan90}. On the contrary, the representation for the heat current has many subtleties and is derived via a continuity equation following Ref.~\onlinecite{Durst09} 
\begin{align}
\hO(r,\tau)=&-\frac{1}{2m}\left[\dpsi^+(r,\tau)\nabla\psi(r,\tau)+ \nabla\psi^+(r,\tau)\dpsi(r,\tau)\right]\label{Eq:CurrentRepsresentation1}
\end{align}
where the dot denotes an~(imaginary) time derivative. Then, the electrical-current-electrical-current correlation function evaluates to~(as vertex corrections are absent in single-band DMFT)
\begin{align}\label{Eq:K0}
K_0=-\frac{ 2 e^2\pi\hbar}{V}\sum_{k}\ \abs{\nabla \varepsilon(k)}^2 \int d\omega\ A^2(k,\omega)\  \frac{\partial f(\omega)}{\partial\omega}%= \text{\big[A$^2$s$^2$m$^{-1}$J$^{-1}$s$^{-1}$\big]}=\text{\big[AV$^{-1}$m$^{-1}$\big]}.
\end{align}
with $V$ the unit cell volume, $\hbar\nabla\varepsilon=\partial\varepsilon/\partial k $ the group velocity of the charge carriers, $A(k,\omega)$ the electron spectral function and $f(\omega)$ the Fermi function. Note that $K_0=\sigma$ has units of a conductivity $[\textrm{A V$^{-1}$ m$^{-1}$}]$. On the other hand, the heat-current-electric-current correlation function reads
\begin{align}\label{Eq:K1}
  K_1=\frac{2 \abs{e}\pi c_{\textrm{eV}} }{V}\sum_{k}\ \abs{\nabla \varepsilon(k)}^2 \int d\omega\ A^2(k,\omega)\ \omega\ \frac{\partial f(\omega)}{\partial\omega}
\end{align}
where $c_{\textrm{eV}}=11600$ K/eV and $K_1$ has units [AJ$^{-1}$Km$^{-1}$]. Combining $K_0$ and $K_1$ leads to the thermopower
\begin{align}\label{Eq:S2}
S=&-\frac{\kboltz \beta}{\abs{e}}\frac{\sum_{k}\ \abs{\nabla \varepsilon(k)}^2 \int d\omega\ A^2(k,\omega)\ \omega\ \frac{\partial f(\omega)}{\partial\omega}}{\sum_{k}\ \abs{\nabla \varepsilon(k)}^2 \int d\omega\ A^2(k,\omega)\ \frac{\partial f(\omega)}{\partial\omega}}.
\end{align}
Note that this result is expressed in the correct units $\text{[V/K]}=\text{[JA$^{-1}$s$^{-1}$K]}$~($\beta$ times the fraction of $K_1$ and $K_0$ remains dimensionless).\\
For the integral in the denominator of Eq.~\eqref{Eq:S2}, i.e. $K_0$, the contributions with respect to $k$ and $\omega$ enter additively, i.e. an asymmetry of $\abs{\nabla \varepsilon(k)}$ with respect to $k$ or $A(k,\omega)$ with respect to $(k,\omega)$ is not important. The integral in the numerator of Eq.~\eqref{Eq:S2}~($K_1$) has an additional $\omega$ and since $df/d\omega$ is negative, $\abs{\nabla \varepsilon(k)}^2 A^2(k,\omega)\omega df/d\omega$ is positive for $\omega<0$~(holes) and negative for $\omega>0$~(electrons)~\cite{quasi}. Consequently, asymmetry of the group velocity $\abs{\nabla \varepsilon(k)}$ as well as the corresponding asymmetry of the spectrum $A(k,\omega)$ are decisive for a high thermopower~\cite{footnote3}. In our model material \Cobaltate\; this is accomplished by a low group velocity $\nabla\varepsilon(k)$ for electrons, whereas holes move much faster through the system. For the band structure this means we have a flat region directly above $\omega>0$ the Fermi level at $\omega=0$ and a steep slope directly below $\omega<0$. This is the feature usually denoted by ``pudding mold''~(cf. Ref.~\onlinecite{Kuroki07}).\\
From Eq.~\eqref{Eq:K0} it becomes clear that $K_0$ is always positive, since $\partial f/\partial \omega$ is negative for all $\omega$ and the other quantities only enter quadratically. At the same time $K_1$ may be either positive or negative depending whether the hole contribution $\omega<0$ or electron contribution $\omega>0$ dominate. This reflects obviously in the sign of the thermopower $S$.\\
\newpage
\section{Structure and spectral properties of sodium cobaltate}\label{Sec:StructureProp}

In 1997, Terasaki and coworkers~\cite{Terasaki97} identified NaCo$_2$O$_4$~($\aquivalent$ Na$_{0.5}$CoO$_2$) as a promising candidate for thermoelectric applications. Other publications confirmed the results of a positive thermopower of the order of ~100 $\mu$V/K even for other sodium contents~\cite{Yakabe98,Lee06,Kaurav09}. With appearance of superconductivity in the H$_2$O intercalated compounds~\cite{Schaak03,Suguira06}, the material \Cobaltate\; once again proved to exhibit surprising features.\\
Recently, Kuroki \emph{et al.}\cite{Kuroki07} argued in favor of band structure properties playing the most important role in determining the thermopower in \Cobaltate. In their study, Kuroki and co-workers used Boltzmann`s equation to compute the Seebeck coefficient $S(T)$. Here, we go beyond that approach and propose an ``ab initio'' method to compute $S$ by means of linear response theory including correlations by DMFT and disorder by CPA~(cf. Sec.~\ref{Sec:LinearResponse} and \ref{Sec:LDA+DMFT}, respectively). As representative compound, we choose a doping of $x_\text{Na}=0.7$, i.e. \cobaltate.

\subsection{Lattice structure}\label{Sec:LatticeStructure}
 \begin{figure}[tp]
  \centering
   \includegraphics[height=6.5 cm]{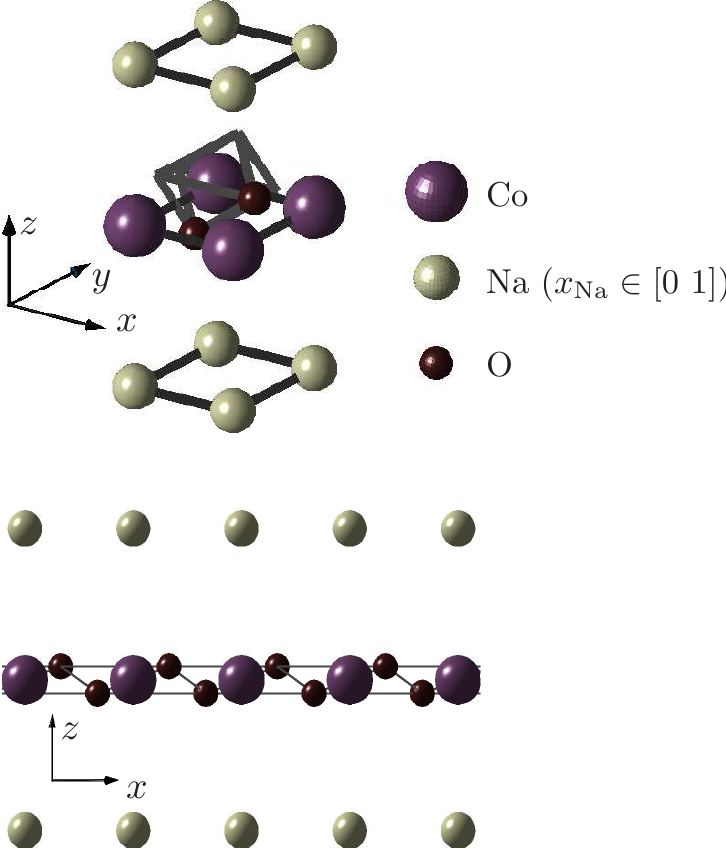}
    \caption{(Color online) Top: Unitcell of NaCoO$_2$ according to the crystal structure from Ref.~\onlinecite{Balsys97}. We additionally visualize the Co and Na bondings as well as one distorted O octahedron. The actual size of the atoms is different from the hard spheres and the distance between the Co and Na layers is larger than shown. For the \cobaltate\; system, we assume that only $70\%$ of the Na position are occupied. Bottom:~The cobalt layers and the distorted oxygen octahedra.}
    \label{Fig:UnitCell1}
\end{figure}
In Fig.~\ref{Fig:UnitCell1}, we plot the crystal structure of the system. Sodium cobaltate NaCoO$_2$ consists of alternating layers of sodium and
cobaltate CoO$_2$. The Co atoms are arranged in a trigonal lattice structure with a nearest-neighbor distance of $2.84\,\AA$, see Ref.~\onlinecite{Balsys97}. Each cobalt atom is surrounded by $6$ oxygen atoms in form of a tilted octahedron. The Na atoms are located in their own layer half way between the CoO$_2$ layers.\\
The exact lateral position of the sodium atoms within their layer and mutual shifts of the CoO$_2$ layers, depend non-trivially on doping, temperature, etc. For example, Huang~\cite{Huang04} reports a change of the Na positions for $\xna=0.75$ around $T=320$ K. On the other hand, an earlier work of Yakabe~\cite{Yakabe98} finds three typical phases of \Cobaltate\; as $\xna$ changes from 0.55 to 1. Since almost all of the transport takes place in the CoO$_2$ layers, mutual shifts of the cobaltate layers only weakly influence the thermopower results. We therefore assume the CoO$_2$ layers to be all equivalent, i.e. there is no structural shift. \\
The lateral position of Na in the compound, on the other hand, is a delicate topic. Many Na coordinates have been reported, mainly the position directly above the Co atom $(x=0,y=0)$ and a position which is aligned to the O atoms $(x=2/3,y=1/3)$~\cite{Balsys97,Singh00}. Moreover, the former reference also suggests movement of the sodium ions. For lower temperatures Zandbergen~\cite{Zandbergen04} found two different types of Co atoms, namely Co$^{3+}$ and Co$^{4+}$. We therefore assume that the cobaltate system shows binary disorder, even if the sodium atoms are distributed on multiple sites. As we are interested in relatively high temperatures for thermoelectric applications, a possible ordering of the Na ions does not appear to be relevant~\cite{footnote1}. For simplicity, we assume the Na to be located on the $(x=0,y=0,z=1/2)$ site directly above the cobalt.  Thus, we can still assign a Na atom to one specific Co site. This corresponds to the same "virtual crystal approximation" which has been used for the underlying band structure calculation~\cite{Singh00}.\\
% \begin{figure}[tp]
%   \centering
%   %  \psfrag{A}{\Huge Co}\psfrag{B}{\LARGE Na}\psfrag{C}{O}
% % \put{500,500}{Co}
%   \begin{picture}(150,150) \thicklines \put(0,0){
% \put(0,0){\epsfig{file=NaCoO2_CrystalStruct2.pdf,height=4.5cm,clip=,angle=0}}
%  % \put(200,-10){\epsfig{file=NaxCoO2_phasediagram_Foo.pdf,clip=,height=6cm,angle=0}}
% \put(96,54){\tiny{$a_{1}$}}
%  \put(106,84){\tiny{$a_{2}$}}
%  }
% 
% \end{picture}
%     \caption{(Color online)~Top view and unit vectors $\va_1$, $\va_2$ on the approximate crystal structure of \Cobaltate~(see Fig.~\ref{Fig:UnitCell1} for the definition of the atoms).}
%     \label{Fig:PhaseDiagram1}
% \end{figure}
\begin{figure}[tp]
 \centering
  \includegraphics[height=5.0 cm]{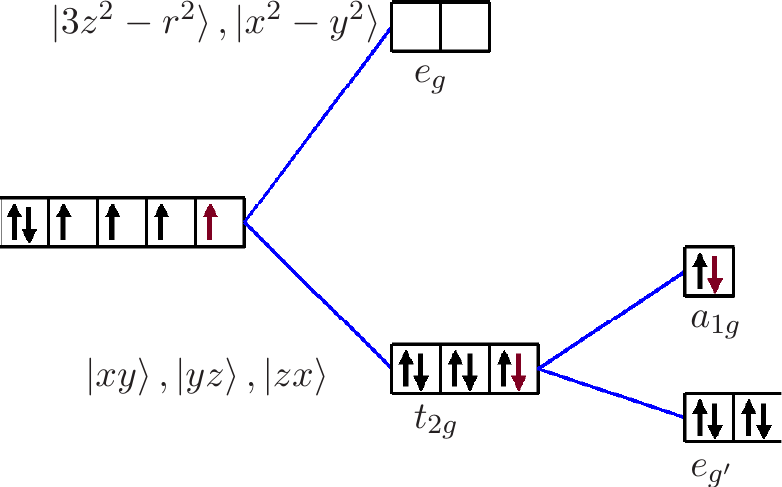}
    \caption{%(left) Visualization of $d$ orbitals. The upper $2$ orbitals are denoted by $e_{g}$, whereas the lower $3$ are denoted by $t_{2g}$.
          (Color online)~The breakup of the degeneracy of the $d$ orbitals: First, the $5$ $d$ basis energetically split up in $3$ $t_{2g}$ orbitals and $2$ $e_g$ orbitals in an octahedral configuration. Second, the $3$ $t_{2g}$ orbitals split up in one $a_{1g}$ and $2$ $e_{g'}$ due to the distortion of the oxygen octahedra. The energy differences shown in this plot are symbolical. We also visualize the occupancy of the orbitals in the different situations, where oxidation state of the cobalt atom in the compound is Co$^{+3}$~(Co$^{+4}$) if a Na partner is (not) present. This additional electron whose presence depends on a sodium donor is shown as an arrow with different color.}
    \label{Fig:CrystalFieldTheory1}
\end{figure}For a doping of $x_\text{Na}=0.7$, 70\% of the sodium sites are actually occupied and we consequently have 30\% vacancies in the sodium layers. The hereby inequivalent Co sites in the cobaltate layers are denoted by \CoNa~($\aquivalent$ Co$^{+3}$) and \CoVac~($\aquivalent$ Co$^{+4}$) depending on whether they have a close sodium atom or not. As we will show below, electrons of one band with little $k_z$ dispersion are responsible for transport. We denote the filling of this band at the two different  sites by \nna\, and \nvac, respectively. Due to the disorder, these electrons tend to be at cobalt sites \CoNa\, with a sodium partner, by means of simple electrostatic attraction to the Na$^{+}$ ionic cores. The strength of this tendency is governed by the disorder potential $\deps$, i.e. the difference of the on-site potential of the two non-equivalent sites. Additionally, many-body effects can influence the movement of electrons and holes in the disordered crystal.\\
To determine the relevant energy bands in this material, it is vital to understand the crystal field splitting of the NaCoO$_2$ lattice. In this light, as part of a thought experiment, the gradual transition between the Co atomic orbitals and their bulk material analogs can be useful: The $5$ atomic \mbox{d-orbitals} can be classified according to their radial and spatial dependence of the corresponding wave functions. In a Cartesian coordinate system one has three planar orbitals $\ket{xy}$, $\ket{yz}$, $\ket{zx}$ denoted by $t_{2g}$ and two orbitals $\ket{3z^2-r^2}$, $\ket{x^2-y^2}$ denoted by $e_{g}$. Refer to Fig.~\ref{Fig:CrystalFieldTheory1} for a visualization of these states. The electron configuration of single-atom cobalt is [Ar]3d$^7$4s$^2$. According to Hund's rule, the \mbox{d-electrons} try to occupy all the orbitals and maximize the total spin~(in this case S=1), if the d-orbitals are degenerate.\\
Atomic orbitals hybridize if the atom is placed in bulk material. However, one can use the picture of atomic orbitals as a starting point to gradually include the influence of the crystal structure. In CoO$_2$, the Co atom is surrounded by an octahedron of $6$ O atoms. The $e_g$ orbitals of Co have a large part of their spatial probability density close to these ligands. Since the O ion is negatively charged, the $e_g$ orbitals pointing towards the oxygen become less favorable for electrons to occupy than in the atomic case. \\
%Additionally to the influence of the octahedral legands, the cobalt orbitals are also shifted by the trigonal arrangement of the cobalt atoms. One of the $e_g$ orbitals is located in the cobalt plane and becomes energetically less favorable. This orbital is denoted by $a_g$, whereas the other two planar orbitals are denoted by $e_{g'}$.
Moreover, the O octahedra which surround the Co atom are distorted, cf. Fig.~\ref{Fig:UnitCell1}. One can show, that this does not lift the degeneracy of the $e_{g}$, but, on the other hand, the $t_{2g}$ states split up in $1$ $a_{1g}$ and $2$ energetically lower $e_{g'}$ orbitals. These $a_{1g}$ and $e_{g'}$ states do not correspond any longer to single $t_{2g}$ orbitals, but are rather linear combinations thereof. These considerations lead to the orbital structure visualized in Fig.~\ref{Fig:CrystalFieldTheory1}. \\
When we go beyond a single Co site, the picture of local orbitals is not fully valid any more, but is rather replaced by bands representing a certain dispersion relation. Still, we can look for bands in the band structure which have predominant $a_{1g}$, $e_{g'}$, or $e_{g}$ orbital character and which arrange in the same energetic order as the orbitals in Fig.~\ref{Fig:CrystalFieldTheory1}.
%\subsection{Band structure}
%\begin{figure}[t]
%  \centering
%  \begin{picture}(400,150) \thicklines \put(0,0){
%  \put(-27,0){\epsfig{file=bandstruct_total.ps,height=5cm,clip=,angle=0}}
%  \put(220,0){\epsfig{file=NaCoO2_Bandstruct_zoomed.pdf,height=5cm,clip=,angle=0}}
  %\put(180,40){$\left\}\rule{0 cm}{0.8 cm}\right.$O p}
  %%\put(180,70){$\left\}\rule{0 cm}{0.2 cm}\right.$ Co $t_{2g}$}
  %\put(180,100){$\left\}\rule{0 cm}{0.2 cm}\right.$ Co $e_{g}$}
  %}
%\end{picture}
%    \caption{Bandstructure plots of NaCo$_2$O$_4$ taken from Singh~\cite{Singh00}. (left) The bands can be energetically separated into three manifolds: First, the oxygen p bands from $-7$ eV to $-2$ eV. Then, the three cobalt d bands with $t_{2g}$ symmetry with an occupation of $n=5.5$ in NaCo$_2$O$_4$. Above the Fermi level, we have the two Co d-e$_g$ orbitals. (right) According to~\cite{Singh00}, the $t_{2g}$ manifold at the Fermi edge split up into two lower e$_{g'}$ band and one a$_{1g}$ orbital, which is half filled for NaCo$_2$O$_4$. The grey area indicates a strong  a$_{1g}$ character.}
%    \label{Fig:Bandstruct1}
%\end{figure}

\begin{figure}[tp]
\includegraphics[height=6cm]{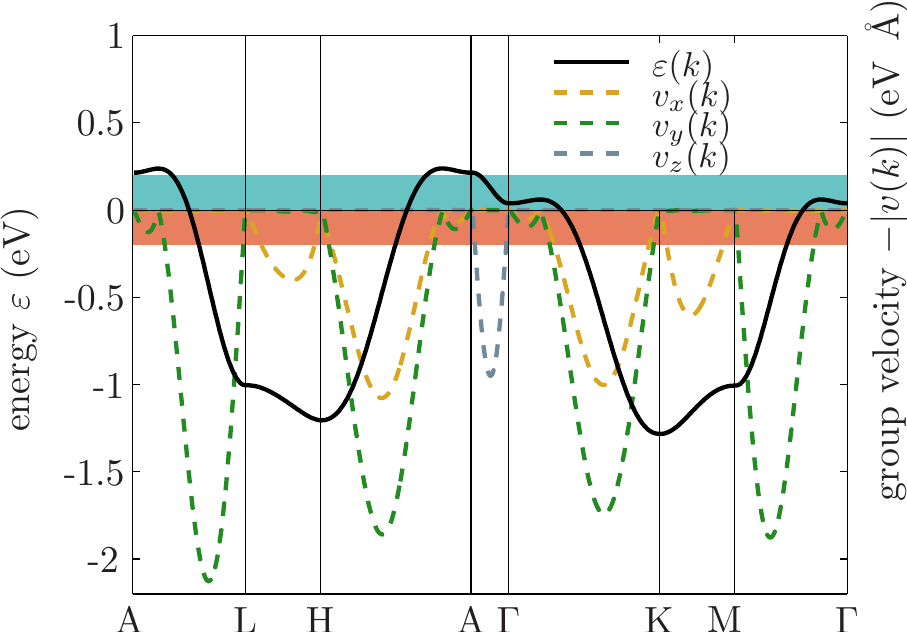}
\caption{(Color online)~Tight-binding approximation~(solid line) and group velocities~(dashed lines) for \cobaltate.}
    \label{Fig:TightBinding1}
\end{figure}
The band structure from Singh~\cite{Singh00} provided the starting point for the tight-binding approximation. To mimic disorder at the level of LDA, Singh used a NaCo$_2$O$_4$ supercell with twice as many bands due to the doubling of the unit cell. %\footnote{This can be seen in Figure~\ref{Fig:Bandstruct2}, which shows two non-degenerate $a_{1g}$ bands.}.
Following a different approach, we extract an approximation for one single a$_{1g}$ band of NaCoO$_2$ from the LDA result~\cite{Singh00} and include the disorder in a second step at the level of the DMFT algorithm, cf. Sec.~\ref{Sec:LDA+DMFT}.\\
Corresponding to atomic valence orbitals, one also finds valence bands within a solid state compound. Analogously, very tightly bound ``core bands'' and, in the ground state unoccupied, ``conductance bands'' may be identified. In cobaltate, the valence bands are the 3 oxygen p-orbitals and the 3 cobalt d-$t_{2g}$-orbitals. Due to high electronegativity, the p-bands of O are well below the Co bands. On the other hand, the sodium s-orbitals belong to the conductance bands and are even above the unoccupied Co $e_g$-bands.
\begin{figure}[tp]
  \centering
\includegraphics[height=6.0cm]{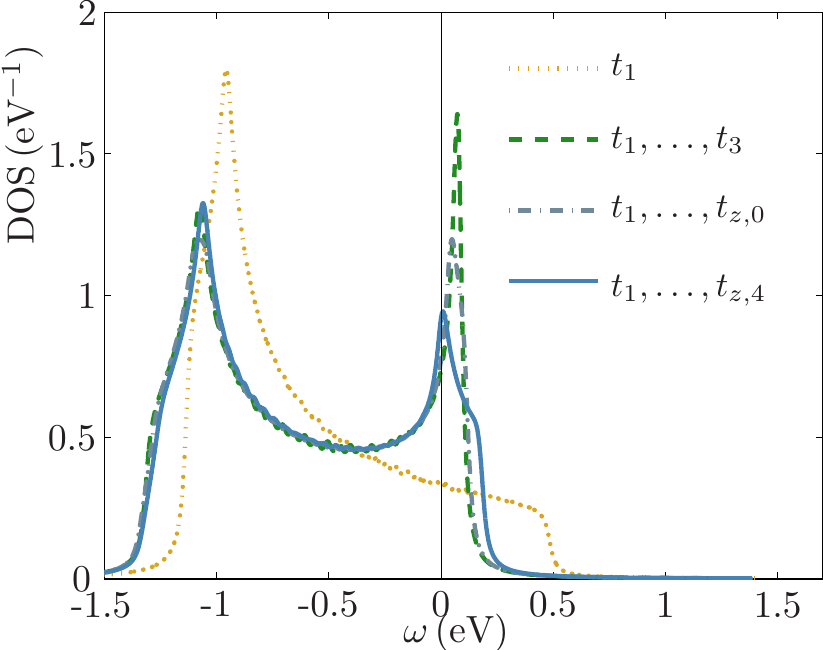}
  \caption{(Color online)~Density of states $N(\omega)$ considering an increasing number of neighbors in the tight-binding approximation using the parameters from Table~\ref{Tab:Tightbinding}. The dotted line corresponds to the DOS computed only with nearest-neighbor hopping, whereas the solid curve is computed with all $9$ parameters. A decisive hopping appears to be the one to the third nearest-neighbor, where a significant change of the characteristics of the DOS can be observed.}
    \label{Fig:DOS1}
\end{figure}
The number of bands considered in the LDA+DMFT model strongly influences the computational effort necessary to solve it. Whereas the decrease to the complexity for the tight-binding approximation is not decisive, the effort for the DMFT calculations heavily depends on the number of bands. We aim therefore to minimize the number of orbitals, without losing~(too much) physically relevant information.\\
The oxidation state of Co in the compound is $+3$ in NaCoO$_2$ and $+4$ in CoO$_2$, respectively. Thus, of the $9$ valence electrons $6$ and $5$, respectively, actually fill the cobalt d bands.
%let us assume that the $2$ $e_g$ states are empty all the time. Analogously, we may argue that the $2$ $e_{g'}$ bands remain constantly filled. All the dynamics of the system is then included by the description of the $a_{1g}$ band, which is filled by an average of $n=1.7$ electrons per site.\\
For transport properties a rather small energy interval of the order of $\OO$($k_\text{B}T$) around the Fermi edge is important, since electrons in this regime can be thermally activated. Only the $t_{2g}$ bands are located within that interval~(under normal conditions) . In the following, we will assume that the hybridization of $e_{g'}$ and $a_{1g}$ bands is negligible. With this assumption, the crystal field splitting visualized in Fig.~\ref{Fig:CrystalFieldTheory1} guarantees that the $e_{g'}$ orbitals remain filled and that, for transport calculations, restricting to the partly occupied $a_{1g}$ orbital is sufficient.\\
To extract the two effective $a_{1g}$ bands from the band structure of NaCo$_2$O$_4$, Kuroki and Usui made a tight-binding fit~\cite{short}, which is shown in Fig.~\ref{Fig:TightBinding1}. Since we aim to include only one a$_{1g}$ band in the DMFT, we have to find the corresponding single tight-binding band for NaCoO$_2$. Though NaCoO$_2$ is similar to NaCo$_2$O$_4$, there are subtle differences: Since the super cell for NaCo$_2$O$_4$ contains two Co atoms the corresponding Brillouin zone has half the size of the one for NaCoO$_2$ in z direction. Thus, the single $a_{1g}$ band will not match exactly one of the two bands for NaCo$_2$O$_4$ of the original fit. The lower band of the two highlighted bands mainly belongs to the cobalt site in NaCo$_2$O$_4$ with a sodium partner. As an approximation, we therefore take the tight-binding parameters for this band, but use the~(larger) Brillouin zone of NaCoO$_2$ in DMFT. The final $a_{1g}$ band is shown in Fig.~\ref{Fig:TightBinding1} together with the corresponding group velocities $\nabla\varepsilon^{\text{TB}}(k)=\{\vv_x,\vv_y,\vv_z\}$.\\
In the following, we discuss the applied tight-binding approximation: Let the $a_{1g}$ band have the dispersion relation $\eLDA(k)$ given by the data points which are the output of the LDA calculation. We now aim to replace $\eLDA(k)$ by an analytic expression, which then can be easily used to obtain the group velocity $\nabla\eLDA(k)$. To this end, we make a tight-binding ansatz with 9 parameters
\begin{align}\label{Eq:tight-binding1}
 \eLDA\big(k\big)\approx \eTB\big(k\big)=\sum_{n=1}^{9}t_n \sum_{j=1}^{\#NN} e^{ik\cdot r_{j_n}},
\end{align}
where \#NN denotes the number of nearest-neighbors of $n$th order and $r_{j_n}$ enumerates the corresponding $n$th nearest-neighbor position. Due to inversion symmetry the exponential function in~\eqref{Eq:tight-binding1} can be simplified as
\begin{align}\label{Eq:tight-binding2}
 \eLDA\big(k\big)\approx \eTB\big(k\big)=2\sum_{n=1}^{9}t_n \sum_{j=1}^{\#\overline{NN}} \cos\big(k\cdot r_{j_n}\big).
\end{align}
Here, $\#\overline{NN}$ means that the pairs of atoms connected by inversion symmetry are counted only once. The first $4$ parameters $t_1,\ldots,t_4$ describe the hopping between nearest-neighbors within one Co-plane. The fifth parameter $t_{z0}$ measures the probability of Co-interplane hopping from a cobalt atom to another cobalt directly above~(below). The final $4$ parameters $t_{z1},\ldots,t_{z4}$ describe processes of combined interplane and nearest-neighbor hopping.\\
The fitted set of parameters~\cite{short} are listed in Table \ref{Tab:Tightbinding}. The final tight-binding fit of the a$_{1g}$ band yields a bandwidth of $1.5$~eV and is, as expected, predominately dispersive in the x-y-plane~(corresponding to the cobaltate layers).\\
\begin{table}[b]
\centering
\begin{tabular}{c|c|c|c}
  \hline
     notation & intra/interlayer&NN&values(par.fit)\\
     \hline
     \hline
     $t_1$ & intralayer & 1st &\phantom{-}0.1800\\
     $t_2$ & intralayer & 2nd &-0.0388\\
     $t_3$ & intralayer & 3rd &-0.0270\\
     $t_4$ & intralayer & 4th &\phantom{-}0.0004\\
     $t_{z,0}$ & interlayer &--& -0.0180\\
     $t_{z,1}$ & interlayer &1st& -0.0049\\
     $t_{z,2}$ & interlayer &2st& -0.0016\\
     $t_{z,3}$ & interlayer &3rd& \phantom{-}0.0011\\
     $t_{z,4}$ & interlayer &4th& \phantom{-}0.0005\\
     \hline
  \end{tabular}
\vspace*{0.3cm}
\caption{Tight-binding parameters obtained by a parameter fit to LDA data.}
\label{Tab:Tightbinding}
\end{table}With the analytic dispersion $\eTB(k)$ from Eq.~\eqref{Eq:tight-binding2}, we may investigate which tight-binding parameters are necessary for the characteristics of the non-interacting LDA density of states
\begin{align}\label{Eq:DOS1}
 N(\omega):=-\frac{1}{\pi}\Im\left[\sum_k\frac{1}{\omega-\eTB(k)+i\delta}\right],
\end{align}
with $\delta\to 0^{+}$. The functions $N(\omega)$ considering hopping up to $1,\ldots,9$ nearest-neighbors are plotted in Fig.~\ref{Fig:DOS1}~(the final solid line corresponds to the LDA-DOS). Consistent with the results from Kuroki~\cite{Kuroki07}, we observe that hopping up to the third nearest-neighbor is necessary to obtain the important upper van-Hove-like peak, which will play a role for the explanation of the large thermopower below. If we also consider interplane hopping~(i.e. $t_{z,i}\neq 0$) the peak above $\omega=0$ is less pronounced. Leaving the 2d regime this is the standard behavior for van-Hove peaks, which are usually damped in 3d with respect to lower dimensions. The final DOS is dominated by two major peaks which are approximately 1~eV apart.

\subsection{LDA+DMFT self energy and spectrum of \cobaltate}\label{Sec:Num Self Energy and Spectrum}
\begin{figure}[tp]
  \centering
\includegraphics[height=10.75cm]{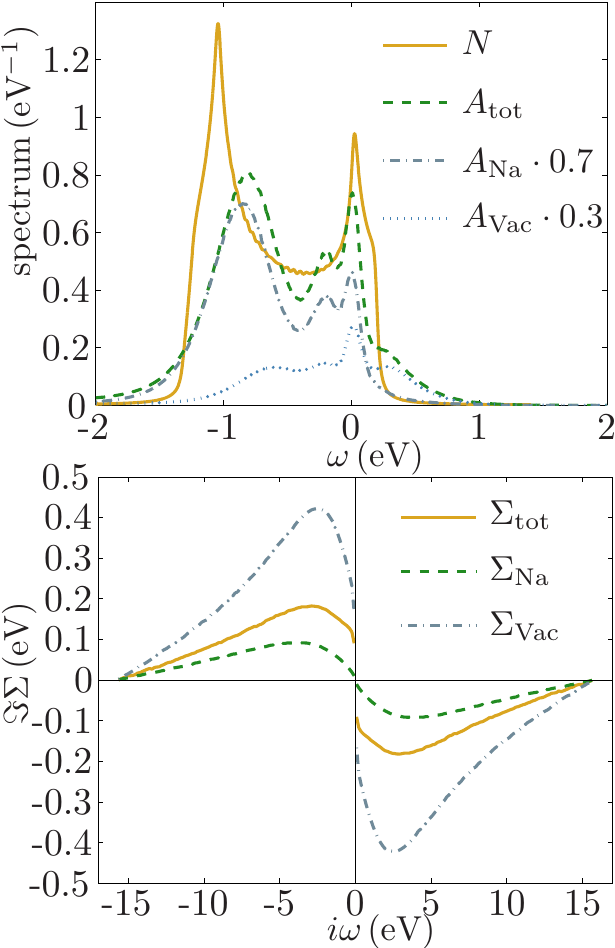}
  \caption{(color online) Top: k-integrated spectra $A(\omega)$ obtained by DMFT and maximum entropy method compared to the non-interacting DOS $N(\omega)$. Bottom: he self energy $\Im\Sigma(i\omega)$ on the imaginary axis. Note that sodium and vacancy contributions are resolved.}
    \label{Fig:SpectrumSelfenergy}
\end{figure}
With the tight-binding dispersion relation $\eTB$, we now run the DMFT algorithm depicted in Sec.~\ref{Sec:LDA+DMFT}. For given temperature $T$, correlation $U$, doping x$_\text{Na}$, and disorder potential $\deps$, we obtain the self energy $\Sigma(i\omega_m)$ at the Matsubara frequencies $\omega_m=(2m+1)\pi/\beta$, and the Greens function $G(\tau)$. Here, and throughout this work, $\beta$ denotes the inverse temperature in eV$^{-1}$ defined by $\beta:=11600/T$~eV$^{-1}$. From the imaginary part of
the self energy $\Im\Sigma(i\omega)$ we can extract information about the strength of
correlation and disorder effects in the system. For example, a divergent
\begin{align}\label{Eq:limitselfenergy1}
\lim_{\omega\rightarrow 0^{\mp}} \Im\Sigma(i\omega)=\pm\infty
\end{align}
indicates that the charge carriers are localized which leads to an insulating phase. On the other hand, a correlated metal usually yields
\begin{align}\label{Eq:limitselfenergy2}
\lim_{\omega\rightarrow 0^{\mp}} \Im\Sigma(i\omega)=0,
\end{align}
with a quasiparticle weight $Z\neq 1$.\\
In Fig.~\ref{Fig:SpectrumSelfenergy}~(bottom), we plot the self energy $\Im\Sigma$ over the Matsubara frequencies $\omega_m$ for our set of initial parameters
\begin{align}\label{Eq:Initial parameters}
\{T=290\ \text{K},\,U=3.5\ \text{eV},\,x_\text{Na}=0.7,\,\deps=0.55\ \text{eV}\}.
\end{align}
 Note that here we employ similar parameters as in a previous LDA+DMFT calculation for \cobaltate~\cite{Kotliar07} which did, however, not analyze transport properties~\cite{footnote2}.
At the first glance, the overall $\Im\Sigma_\text{tot}$ shows characteristics of a bad metal~\cite{ulmke}, since in the whole temperature range considered it neither diverges nor converges to $0$ for $\omega\rightarrow 0$. Hence, the separation of correlation and disorder effects on $\Im\Sigma$ needs further investigation. As a first step, we can also visualize the contributions of the two sites \CoNa\, and \CoVac\, to the self energy $\Im\Sigma_\text{tot}$. Following Sec.~\ref{Sec:LDA+DMFT}, these contributions are computed via Dyson's equation with the corresponding Greens function, i.e.
\begin{align}\label{Eq:SelfEnergy_sep1}
  \Sigma_\text{Na} =&\quad \GG_0^{-1}+\deps-G^{-1}_\text{Na},\\
  \Sigma_\text{Vac} =&\quad \GG_0^{-1}-G^{-1}_\text{Vac},\\
  \Sigma_\text{tot} =&\quad \GG_0^{-1}-G^{-1},
\end{align}
where $G=\xna G_\text{Na}+\xvac G_\text{Vac}$. We observe that the self energy $\Im\Sigma_\text{Na}$ for the sites \CoNa\, is metallic, the other contribution $\Im\Sigma_\text{Vac}$ for \CoNa\, showing insulating tendency. \\
Aquivalent to the self energy $\Im\Sigma(i\omega)$ connected to the influence of correlation \emph{and} disorder, the Greens function $G(\tau)$ provides information about the many-body excitations of the system, as the k-integrated spectrum  $A(\omega)$ is given implicitly by

\begin{figure}[tp]
  \centering
 \includegraphics[height=5.75cm]{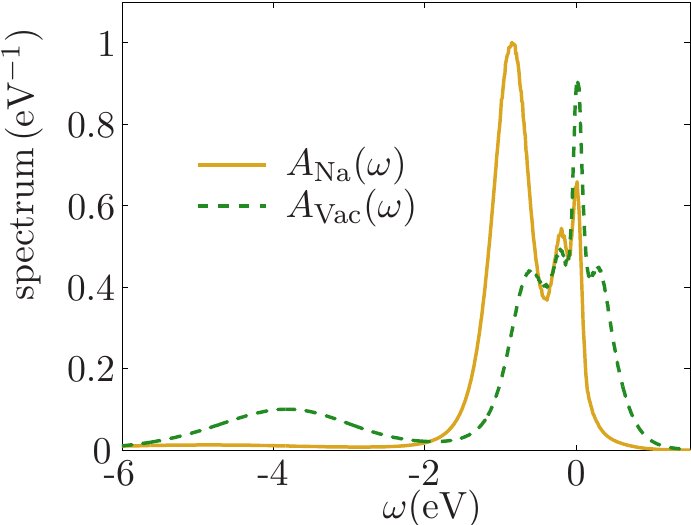}
    \caption{(Color online) The two contributions $A_\text{Na}(\omega)$ and $A_\text{Vac}(\omega)$ to the overall spectrum $A_\text{tot}(\omega)$ \emph{without} the stoichiometric weighting factor. The filling at the sites \CoNa\, is \nna$=0.91$, therefore correlation effects are negligible. For \CoVac\, with \nvac$=0.71$  on the other hand, the formation of a quasiparticle peak at the Fermi edge $\omega=0$, a lower Hubbard band at $\omega\approx-U=-3.5$ eV and a upper Hubbard band at $\omega\approx0.5$ eV can be observed. The center of the two spectra are approximately shifted by the disorder potential $\deps=0.55$ eV with respect to each other.}
    \label{Fig:Spectrum2}
\end{figure}
\begin{figure}[tp]
  \centering
 \includegraphics[height=6.0 cm]{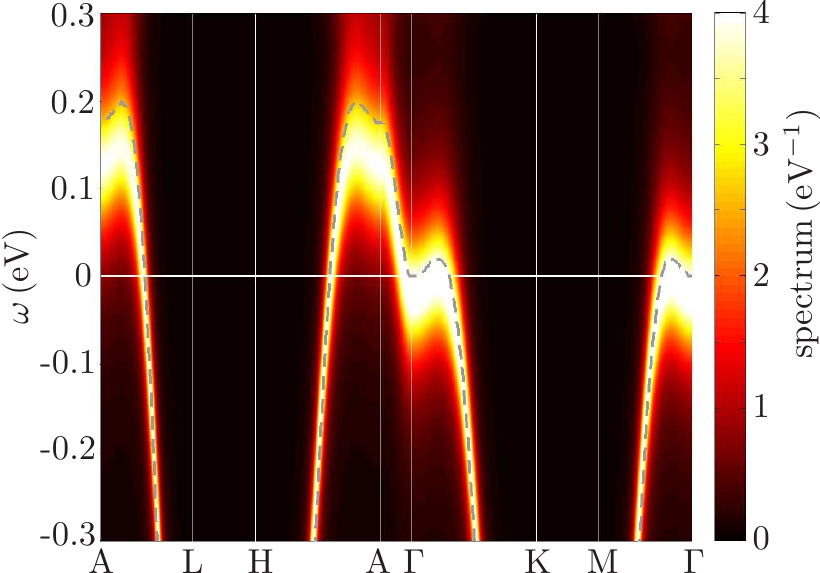}
    \caption{(Color online) Spectral function $A(k,\omega)$ for $T=290$ K obtained by a Taylor fit of the DMFT self energy. The color code of the values of $A(k,\omega)$ is given by the colorbar on the right-hand side. The dashed line corresponds to the non-interacting tight-binding dispersion relation.}
    \label{Fig:Spectrum1}
\end{figure}
\begin{align*}
G(\tau)=\int d\omega\frac{e^{\tau(\mu-\omega)}}{1+e^{\beta(\mu-\omega)}}\ A(\omega).
\end{align*}
To calculate the spectral functions, we employ the maximum entropy method~(see for example Ref.~\onlinecite{Skilling89,Avella07}). In Fig.~\ref{Fig:SpectrumSelfenergy}~(top), we plot the total spectrum $A_\text{tot}(\omega)$ for the a$_{1g}$ band and compare it to the contribution of the two sites \CoNa,\CoVac\, as well as to the non-interacting density of states~(DOS) $N(\omega)$. Note that we do not show here the lower Hubbard band at $\omega\approx-U=-3.5$eV. At the first glance, the changes between the $N(\omega)$ to $A_\text{tot}(\omega)$ are not significant. But since two contributions $A_\text{Na}$, $A_\text{vac}$ with two different electron fillings
%This is not surprising, due to the fact that the overall filling \ntot=N/2=0.85 is distributed as
\begin{align}\label{Eq:filling1}
n_\text{Na}=2\cdot0.91,n_\text{Vac}=2\cdot 0.71\quad(n_\text{tot}=2\cdot0.85=1+\xna)
\end{align}
mix, a separate investigation of the two spectra is useful~(note that $n_\text{Na}\xna+n_\text{Vac}\xvac=n_\text{tot}$). On \CoNa\; sites, the a$_{1g}$ band is almost filled, whereas on \CoVac\, the value is closer to half filling, where correlation effects are expected to be stronger. In fact, as can be seen in Fig.~\ref{Fig:Spectrum2}, the formation of a quasiparticle peak at $\omega=0$ and a lower Hubbard band at $\omega\approx -U=-3.5$ eV in the spectrum $A_\text{Vac}$ for the sites \CoVac\, can be observed. Especially the Fermi liquid behavior for small $\omega$, may be a decisive ingredient for higher thermopower, since the spectral weight in the thermally activated energy interval is crucial as already discussed in Sec.~\ref{Sec:LinearResponse}. This analysis indicates that disorder enhances the effect of correlations in the compound driving the sites Co with Na vacancy, closer to half filling. As a consequence, the quasiparticle peak for a given correlation~$U$ is more pronounced than in the case without disorder.\\
Analogously to the filling, one can also investigate the approximate renormalization of the two bands
\begin{align}\label{Eq:Renorm1}
Z_\text{Na}=0.88,Z_\text{Vac}=0.36\quad(Z_\text{tot}=0.48),
\end{align}
which was obtained from~$\Im \Sigma(i\omega_m)$ by using the tangent of the two Matsubara frequencies closest to the origin. This values should be considered with caution, since the disorder perturbs the Fermi liquid behavior at both sites \CoNa\, and \CoVac. However, as Fig.~\ref{Fig:Spectrum2} indicates, \cobaltate\; can be interpreted as a system consisting of sites with large correlations \CoVac, and sites \CoNa\, where correlation effects are weak.\\
Apart from the k-integrated spectrum $A(\omega)$, we are also interested in the k-resolved quantity $A(k,\omega)$, that is inserted in Kubo's formula~\eqref{Eq:S2}. Therefore, we need an analytic continuation $\Sigma(i\omega)\rightarrow \Sigma(\omega)$, which is, in general, non-trivial. The behavior of $\Sigma(\omega)$ further away from the Fermi level is not important for transport properties and a Taylor fit is hence sufficient.
%However, a better approximation for the total $\Sigma(\omega)$ for larger $\omega$, can be obtained for instance by applying a fit to the Greens function $G_\text{R}$ on the real axis, computed from the k-integrated spectrum $A(\omega)$. The total workflow reads
%\begin{align}\label{Eq:Workflow1}
% \text{DMFT}\longrightarrow G(\tau)\text{(imag. axis)}\overset{\text{Max. Ent}}{\longrightarrow}A(\omega)\overset{\text{Kramers-Kronig}}{\longrightarrow}G_\text{R}(\omega)\text{(real axis)}\overset{\text{fit}}{\longrightarrow}\Sigma(\omega).
%\end{align}
With the self energy $\Sigma(\omega)$ the spectrum is calculated as
\begin{align}\label{Eq:SpectrumfromSelfenergy}
 A(k,\omega)=-\frac{1}{\pi}\frac{\Im(\Sigma(k,\omega))}{\left[ \omega +\mu-\varepsilon(k)-\Re(\Sigma(k,\omega))\right]^2+\left[\Im(\Sigma(k,\omega))\right]^2},
\end{align}
where $\mu$ is the chemical potential from the DMFT calculation. \\
A visualization of the approximation for $A(k,\omega)$ compared to the tight-binding band can be found in Fig.~\ref{Fig:Spectrum1}.
%Between the energies $-0.7$\,eV to $-0.1$\,eV the results from the Maximum entropy method were numerically unstable and have thus been replaced by those of the smooth Taylor fit. Though the renormalization of the band appears to be small, correlation effects can still remain important for the thermopower as we show below.

%\subsection{Comparison of the spectrum to experiment}\label{Sec:Comparison to experiment}
\begin{figure}[tp]
  \centering
\includegraphics[height=9.0 cm]{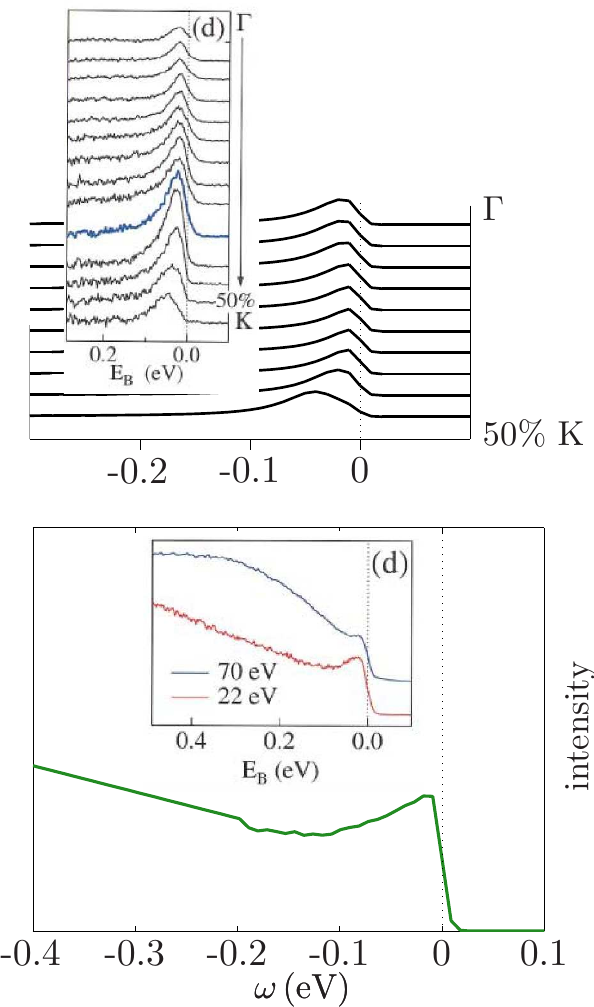}
    \caption{(color online) Top: Comparison of the spectrum $A(k,\omega)$ to ARPES results, inset taken from Ref.~\onlinecite{Yang04}). In experiment, the peaks correspond to a$_{1g}$ excitations. Bottom:~k-integrated spectrum $A(\omega)$ and corresponding experimental results~(inset taken from Ref.~\onlinecite{Yang04}). The two slopes in the inset are spectra for different photon energies. For comparison, a linear background signal for $\omega<-0.2$ eV has been added by hand to the numerical results.}
    \label{Fig:Experiment1}
\end{figure}
%In the presented approach, the band structure of \cobaltate\, has been approximated by a single-band tight-binding fit in order to compute transport properties. 
Let us remark here that it cannot be expected that the electronic structure or spectra match the experimental results in a large energy interval around $\omega=0$ for a model without the e$_{g'}$ bands. However, ARPES data from Yang~\cite{Yang04} indicates that the e$_{g'}$ bands, which appear to cross the Fermi edge, actually do not contribute to the Fermi surface, i.e. are shifted down in energy by many-body effects. Thus, a comparison to experimental data may well be accurate in the vicinity of the Fermi level.\\
In Fig.~\ref{Fig:Experiment1}, we show the numerically obtained spectra and the corresponding ARPES data~\cite{Yang04}. For a direct  comparison, all numerical results are already multiplied with the Fermi function $f(\omega,\beta)=(e^{\beta\omega+1})^{-1}$ with $T=290$K as in the experiment~(we thus assume that the spectra do not change qualitatively as $T_\text{num}=290$ K$\rightarrow T_\text{exp}=40$\,K). On the left-hand side the k-resolved spectrum $A(k,\omega)$ is plotted for various values of $k$ along the $k$-path $\Gamma\rightarrow K$. In the inset, we give the corresponding experimental results. The reason for the good agreement is the large a$_{1g}$ character of the experimental spectra. On the right-hand side of Fig.~\ref{Fig:Experiment1}, we compare the k-integrated spectrum $A(\omega)$. To simulate the e$_{g'}$ background we add a linear function for $\omega<0.2$ eV. Again the data coincide qualitatively, as the inset with the experimental results shows.

\newpage
\section{Thermopower}\label{Sec:Thermopower}
\begin{figure*}[p]
  \centering
\includegraphics[height=20.0 cm]{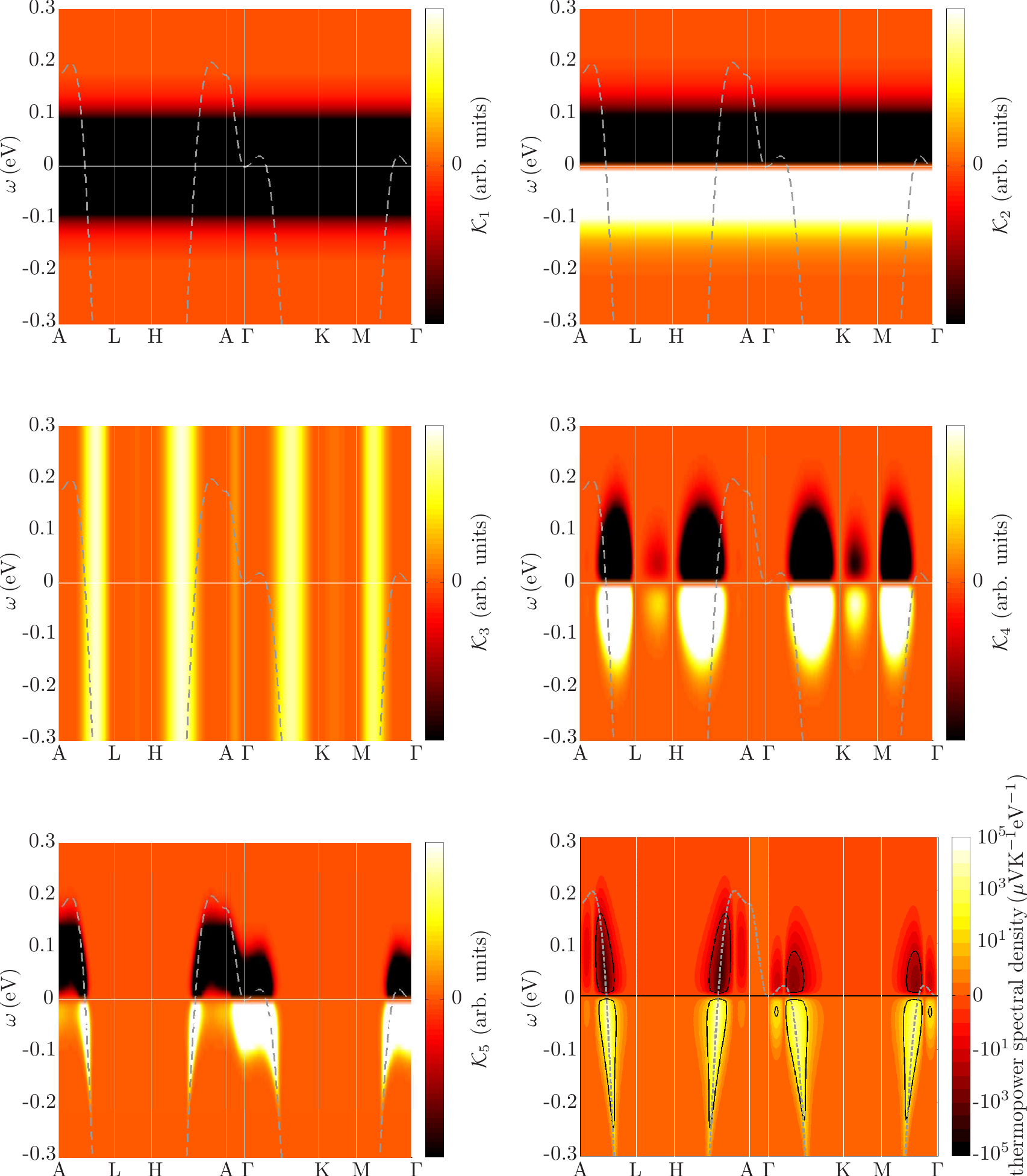}
    \caption{(color online) Contributions $\KK_1,\ldots,\KK_5,\KK_\text{tot}$ to the thermopower $S$ for T=290 K according to Equations~\eqref{Eq:KubosformulainResults Kernels1}-\eqref{Eq:KubosformulainResults Kernelstot}. Positive values are emphasized by bright~(yellow) domains, negative contributions are visualized by dark~(red) domains. The units of the kernels $\KK_1,\ldots,\KK_5$ are arbitrary, but domains with brighter or darker scale~(intenser red or blue colors) correspond to larger positive or negative values, respectively. As a dashed line, we also plot the LDA band structure. The solid line shows the isovalues $\pm 10^2$ $\mu$VK$^{-1}$eV$^{-1}$ for the thermopower spectral density.}
    \label{Fig:Spectralcontributions1}
\end{figure*}

In order to understand the influence of temperature, disorder, correlations and doping on the self energy and the thermopower, in this section we will investigate their effect separately with the other parameters kept fixed. For disorder and correlations, however, it is revealed that their effect strongly depends on the respective other parameter. We therefore discuss combined effects of disorder and correlations in an own subsection. Before starting with the first parameter, let us introduce some helpful quantities which will be helpful in the analysis.
In Sec.~\ref{Sec:LinearResponse}, we derived the expression $S(T)\sim K_1/K_0$ for the thermopower. The pivotal quantity which we now focus on, is $K_1$ depicted in Eq.~(\ref{Eq:K1}) 
% \begin{align}\label{Eq:KubosformulainResults}
%   K_1(T)=11600\sum_k \int d\omega\ \abs{\nabla \varepsilon^{\text{TB}}(k)}^2\ A^2(k,\omega,T)\ \frac{\partial f(\omega,T)}{\partial \omega}\ \omega,
% \end{align}
and therein especially the kernel as a function of $(k,\omega)$. In most of the cases, changes of $K_1$ directly reflect on the thermopower $S$.
If we separate the factors in the kernel for the initial parameters~\eqref{Eq:Initial parameters}, we arrive at contributions visualized in Fig.~\ref{Fig:Spectralcontributions1}. Investigating these contributions can lead to a deeper understanding of the important processes in thermoelectric transport. In order to factorize the total kernel $\KK_\text{tot}$, we define $5$ auxiliary functions of $k$ and $\omega$
\begin{align}
  \KK_1(k,\omega)&:=\frac{\partial f(\omega,T)}{\partial \omega},\label{Eq:KubosformulainResults Kernels1}\\
  \KK_2(k,\omega)&:=\frac{\partial f(\omega,T)}{\partial \omega}\ \omega,\label{Eq:KubosformulainResults Kernels2}\\
  \KK_3(k,\omega)&:=\abs{\nabla\varepsilon(k)}^2,\label{Eq:KubosformulainResults Kernels3}\\
  \KK_4(k,\omega)&:=\abs{\nabla\varepsilon(k)}^2\frac{\partial f(\omega,T)}{\partial \omega}\ \omega\label{Eq:KubosformulainResults Kernels4}\\
  \KK_5(k,\omega)&:=A^2(k,\omega)\frac{\partial f(\omega,T)}{\partial \omega}\ \omega,\label{Eq:KubosformulainResults Kernels5}\\
  \KK_\text{tot}(k,\omega)&:= \frac{\kboltz}{T K_0}\frac{2\pi e}{V}\abs{\nabla\varepsilon(k)}^2\ A^2(k,\omega)\frac{\partial f(\omega,T)}{\partial \omega}\ \omega.\label{Eq:KubosformulainResults Kernelstot}
\end{align}
where we included all prefactors in $\KK_\text{tot}$ such that $S=\sum_k\int d\omega \KK_\text{tot}(k,\omega)$. The total kernel $\KK_\text{tot}$ may thus be interpreted as a ``thermopower spectral density'' determining the contribution to $S$ in $(k,\omega)$-space. In the following we restrict ourselves to in-plane transport, i.e. we set $\abs{\nabla\varepsilon(k)}^2=\vv^2=\vv_x^2+\vv_y^2$.\\
In Fig.~\ref{Fig:Spectralcontributions1}, a visualization of $\KK_1,\ldots,\KK_5,\KK_\text{tot}$ for \{$T=290$ K, $U=3.5$ eV, $\deps=0.55$ eV\} is shown.
The derivative of the Fermi function $\KK_1$ constitutes an energy interval around the Fermi edge, where electrons are thermally activated. There will be no electronic transport outside of this interval. The different sign of the charge carriers electrons~(negative) and holes~(positive) is accounted for by the factor $\omega$ in $\KK_2$. Thus, Fig.~\ref{Fig:Spectralcontributions1}(top,right) shows the charge carriers activated for transport with respect to sign. On the other hand, the group velocity $\KK_3$ is a quantity which only depends on the band structure. As a consequence, $\KK_3$ will be large whenever the slope of $\gv(k)$ is steep, cf. Fig.~\ref{Fig:Spectralcontributions1}~(middle,left). Combining the latter three contributions leads to $\KK_4$ shown in Fig.~\ref{Fig:Spectralcontributions1}~(middle,right). We observe that the values are still distributed symmetrically with respect to the Fermi level. The effect of the temperature and spectrum without the group velocity is encoded in $\KK_5$. There appear to be more contributions from the electrons $\omega>0$~(Fig.~\ref{Fig:Spectralcontributions1}~(bottom,left)).
Finally, the thermopower spectral density $\KK_\text{tot}$ indicates the regions of $(k,\omega)$ space which contribute to the thermopower, cf. Fig.~\ref{Fig:Spectralcontributions1}~(bottom,right). The function $\KK_\text{tot}$ yields both positive~(holes) and negative~(electrons) contributions to the thermopower, which may in principle also annihilate each other. In fact, that is the usual behavior for most materials where the thermopower $S$ is therefore only of the order of $\pm\OO(1)\mu$V/K. In our example, the electron contributions are significantly smaller than hole contributions, and this inbalance reflects itself on a positive thermopower $S$.

\subsection{Effects of the temperature}\label{Sec:Temperature Effects}
% \begin{figure}[t]
%   \centering
%   \begin{picture}(150,200) \thicklines \put(0,5){
%   \put(0,100){\epsfig{file=dfdwtimesw1.pdf,height=4.75cm,clip=,angle=0}}
% \put(0,0){\epsfig{file=dfdwtimesw2.pdf,height=3.75cm,clip=,angle=0}}
% \put(-40,100){\rotatebox{90}{{\small$\omega$~[eV]}}}
% \put(70,-10){\small$T$~[K]}
% %right
% \put(198,100){\rotatebox{90}{{\small$\omega\,$[eV]}}}
% \put(300,-10){\small$T$~[K]}
% \put(330,51){\small$\Delta$ T}
% \put(368,82){\small$\frac{\ln 2}{11600}$ $\Delta$ T}
%   }
% \end{picture}
% \caption{(left) The dimensionless kernel $\KK_2=(\partial f(\omega,\beta)/\partial \omega)\ \omega$ from Equation~\eqref{Eq:KubosformulainResults Kernels2} as a function of $(T,\omega)$. Large intensity of the red and  correspond to large positive and negative values of $\KK_2$, respectively. (right)~ The integral $\int_{-\infty}^0 d\omega\ \KK_2(\omega)$ in eV over temperature~$T[K]$. A linear growth with a slope of $\ln(2)/11600$ can be observed.}
%     \label{Fig:temperatureEffects1}
% \end{figure}
\begin{figure}[t]
  \centering
 \includegraphics[height=14.0 cm]{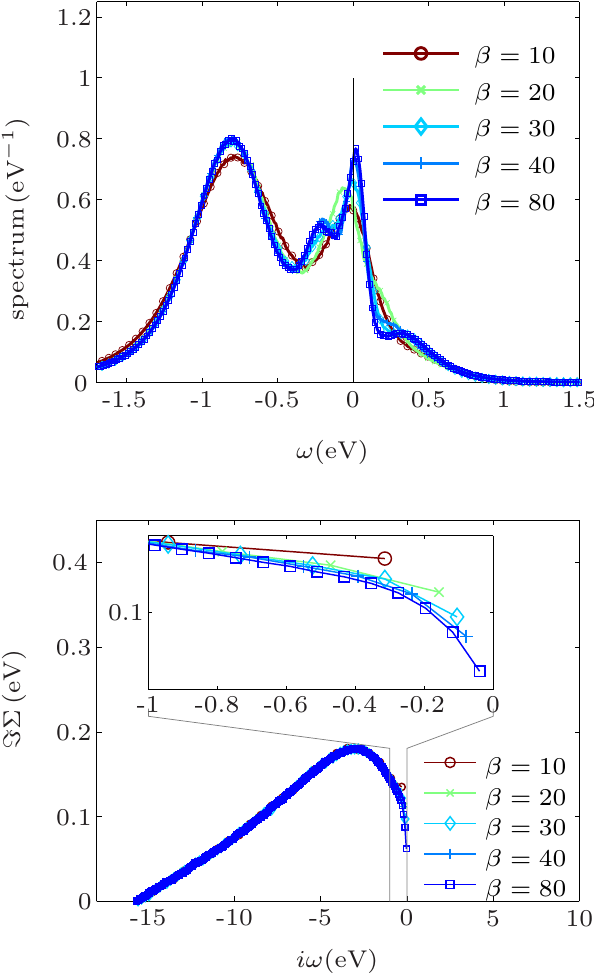}
    \caption{(color online) Top: k-integrated spectra $A(\omega)$ for various temperatures $\beta$ in eV$^{-1}$. The spectral weight around the Fermi edge is decreasing with respect to temperature $T$. There is no major change of $\Im\Sigma(i\omega)$ with respect to temperature, except for Matsubara frequencies $\omega_m$ close to the $\omega=0$. Bottom:~Self energy $\Im\Sigma(i\omega)$ on the Matsubara axis for various temperatures.}
    \label{Fig:selfenergy+spectrum_temperature}
\end{figure}

\begin{figure*}[p]%Akw_N1.7_B40_U3.5_deps0.55.pdf
  \centering
 \includegraphics[height=20.5 cm]{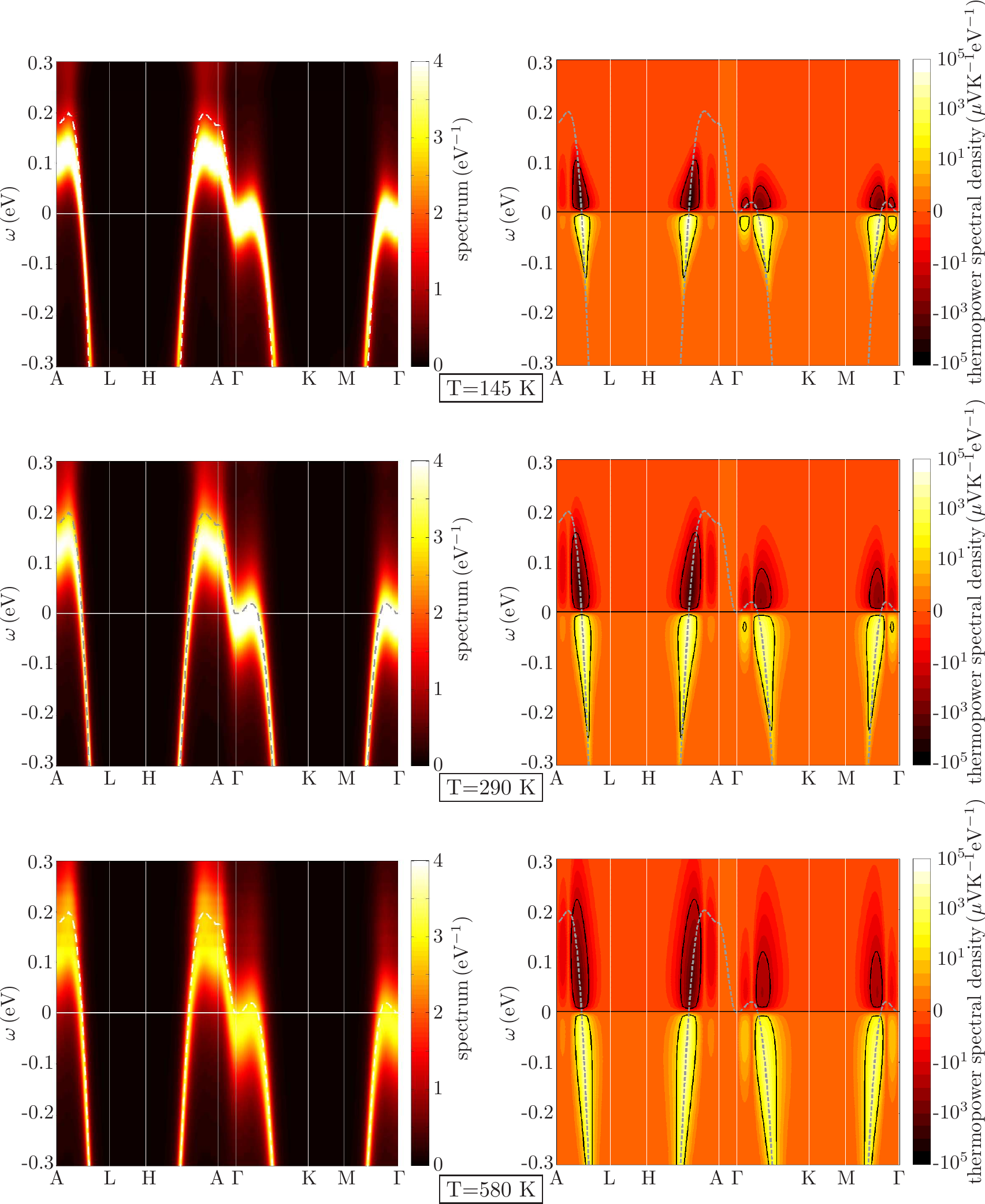}
    \caption{(color online) k-resolved spectra $A(k,\omega)$~(left) and corresponding thermopower spectral densities \hspace*{0.05cm}  $\KK_\text{tot}$~(right) for various temperatures $T$. The color code is given by the colorbar. The dashed line on the spectral images is the tight-binding fit. The spectral weight around the Fermi edge is decreasing with temperature. Thus, the increase of the contribution to the thermopower gets dampened and is therefore not linear.}
    \label{Fig:Spectralcontributions_temperature}
\end{figure*}

Before considering the numerical results of the model system \cobaltate, we discuss the effects of temperature to the thermopower $S$ by a rather general approach analyzing the thermal dependence of the kernel $\KK_5(k,\omega)$.
% \begin{align}\label{Eq:TemperatureEffects_Kernel1}
% =A^2(k,\omega,T)\ \frac{\partial f(\omega,T)}{\partial \omega}\ \omega.
% \end{align}
Let us assume for the moment the extreme case that the spectrum $A(k,\omega)\approx A(k)\Theta(-\omega)$, i.e. only holes contribute to the thermopower. Then, we can solve the integral
\begin{align}\label{Eq:TemperatureEffects_Kernel2}
 I(\beta):=\int_{-\infty}^{0} d\omega\ \frac{\partial f(\omega,\beta)}{\partial \omega}\ \omega
\end{align}
to obtain an approximation of the thermal dependence of the thermopower $S$:%, cf. Figure~\ref{Fig:temperatureEffects1}~(left) for a visualization of the kernel.
% By partial integration, we find
\begin{align*}%\label{Eq:TemperatureEffects_Kernel1}
 I(\beta)&=\left.f(\omega,\beta)\ \omega\ \right|_{-\infty}^{0}-\int_{-\infty}^{0}d\omega\, f(\omega,\beta) \\
% &=\left.\frac{\omega}{1+e^{\beta\omega}}\right|_{-\infty}^{0}-\left[\omega-\frac{1}{\beta}\ln(1+e^{\beta\omega})\right]_{-\infty}^{0}\\
%&=\frac{\ln(2)}{\beta}+\lim_{w_0\rightarrow -\infty}\Big[-\underbrace{\frac{\omega_0}{1+e^{\beta\omega_0}}}_{\rightarrow \omega_0}+\omega_0-\underbrace{\frac{1}{\beta}\ln(1+e^{\beta\omega_0})}_{\rightarrow 0}\Big]\\
&=\frac{\ln(2)}{11600}\ T\ \text{[eV]}.
\end{align*}
Analogous considerations for the conductivity $\sigma=K_0$ yield a factor $\beta$ which cancels with $\beta$ from the definition of $S=\kboltz\beta e^{-1} K_1 K_0^{-1}$. Thus, a linear dependence $S\propto T$ represents a first approximation for the thermopower. Since extreme asymmetries as assumed in Eq.~\eqref{Eq:TemperatureEffects_Kernel2} will typically not persist for larger $T$  the linear dependence will be reduced for higher temperatures.\\
For low temperatures $T$, the spectrum $A$ inside of the thermally activated energy interval will become symmetric with respect to $\omega$, i.e. $A(k,\omega)\approx A(k)$. Then, the integral
\begin{align*}%\label{Eq:TemperatureEffects_Kernel1}
 I(\beta)=\int_{-\infty}^{\infty}d\omega\ \frac{\partial f(\omega,\beta)}{\partial \omega}\ \omega=0
\end{align*}
vanishes, since the integrand is an odd function. Thus, for small $T\rightarrow 0$, the thermopower $S$ is expected to converge to $0$.\\
After these general considerations, we now return to \cobaltate: In Fig.~\ref{Fig:selfenergy+spectrum_temperature}, we plot the imaginary part of the self energies $\Im\Sigma(i\omega)$ on the Matsubara axis and the k-integrated spectra $A(\omega)$ for various temperatures $T$. The high energy tail of $\Im\Sigma(i\omega)$ is determined by a sum rule~\cite{sumrule1}, and is therefore almost temperature independent. In the vicinity of $\omega=0$, we observe a variation of $\Im\Sigma(i\omega)$ with respect to temperature. This has several reasons: First, the highest Matsubara $\omega_m$ moves closer to $\omega=0$ by definition and therefore we obtain a finer resolution. On the other hand, incoherent effects due to electron-electron scattering decrease with lower temperature.\\
The k-integrated spectrum plotted in Fig.~\ref{Fig:selfenergy+spectrum_temperature}(top), shows less spectral weight of $A(\omega)$ in the vicinity of the Fermi edge for increasing temperature $T$. This is a usual effect of higher temperatures: the dampening of narrow peaks in the spectrum. In addition, there are no indications for a phase change to an insulator of the material in the investigated temperature interval 145--1160 K originated in the electronic structure, since the qualitative spectral distribution remains more or less unchanged. However, the result for $\beta=20$\,[eV$^{-1}$] is puzzling, because it shows a deviation of the general temperature behavior with respect to the other results. The same behavior is also observed for other doping $\xna=0.6,0.8$. Most probably this result is connected to the numerical error in the employed Maximum Entropy Method~(MEM) as two spectral peaks merge into one.\\
%Decreasing spectral weight $A(\omega)$ with respect to temperature $T$ for small $\omega$ influence the numerical result for the thermopower $S$ via Equation~\eqref{Eq:KubosformulainResults}. In order to confirm this suspicion,
In Fig.~\ref{Fig:Spectralcontributions_temperature}, we show how the k-resolved spectra and the thermopower spectral density $\KK_\text{tot}$ changes with temperature. We observe that the spectral weight is in fact decreasing in the thermally activated energy interval. Thus, the linear approximation for $S(T)$ has to be adapted for the spectra of \cobaltate. The linear increase will be dampened with increasing temperature leading to a flattening of $S(T)$. This can also be seen in the final results in Fig.~\ref{Fig:Seebeck1_x}, where we plot the thermopower $S(T)$ for various doping $\xna$.


\begin{figure}[tp]
  \centering
  \includegraphics[height=6.5 cm]{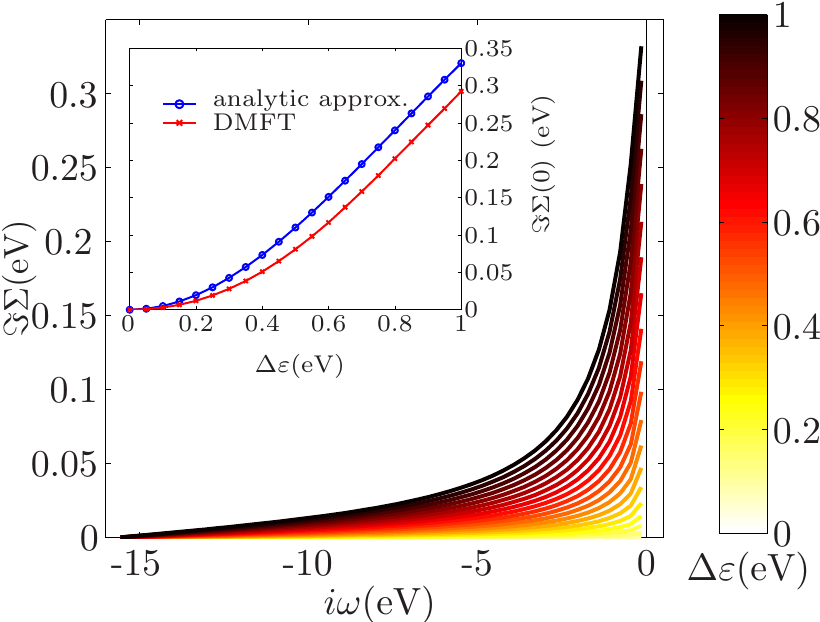}
    \caption{(color online) The imaginary part of the self energy $\Im\Sigma(i\omega)$ on the Matsubara axis for $T=290$ K and $U=0$ eV for various values of the disorder potential $\deps=0\rightarrow 1\,$eV. A gradual increase of $\Im\Sigma(i\omega)$ with respect to $\deps$ is observed. (inset)~The values $\lim_{\omega\rightarrow 0}\Im\Sigma(i\omega)$ obtained by the approximate formula~\eqref{Eq:CPA in results5} and by DMFT. For small disorder $\deps$ the approximation is in good agreement with the DMFT data.}
    \label{Fig:selfenergyandapproximativeformula1 T=290}
\end{figure}

\subsection{Effects of the disorder}\label{Sec:Disorder Effects}
% \begin{figure}[t]
%   \centering
%   \begin{picture}(150,200) \thicklines \put(0,5){
% \put(-42,60){\rotatebox{90}{{ \small DOS\,[eV$^{-1}$]}}}
% %\put(180,-10){$i\omega$[eV]}
% \put(0,120){\epsfig{file=ToyDisorder2.pdf,height=4.2cm,clip=,angle=0}}
% \put(0,70){\epsfig{file=ToyDisorder3.pdf,height=4.13cm,clip=,angle=0}}
% \put(0,0){\epsfig{file=ToyDisorder4.pdf,height=4.2cm,clip=,angle=0}}
% \put(38,-8){\small$\omega$[eV]}
% \put(180,-8){\small$\omega$[eV]}
% \put(325,-8){\small$\omega$[eV]}
% \put(36,123){$\deps=0$}
% \put(156,123){$\deps\approx t/3\approx 0.5$ eV}
% \put(303,123){$\deps=2$ eV$\approx 4t/3$}
% \put(15,95){\small$N_\text{Na}$}
% \put(9,15){\small$N_\text{Vac}$}
% \put(159,95){\small$N_\text{Na}$}
% \put(173,15){\small$N_\text{Vac}$}
% \put(300,95){\small$N_\text{Na}$}
% \put(360,15){\small$N_\text{Vac}$}
%
%   }
% \end{picture}
%     \caption{Qualitative effect of disorder on the~(non-interacting) density of states $N_\text{Na}$, $N_\text{Vac}$ of the sites \CoNa\, and \CoVac, respectively. The bandwidth of \Cobaltate\, is $t\approx 1.5$ eV. At a certain disorder potential $0<\deps<t$ the DOS $N_\text{Vac}$ at the sites \CoVac\, is close to half filling, which is important for correlation effects~(which are not shown in this set of images). For larger potential $\deps>t$, the system becomes insulating -- a result also denoted by alloy band splitting. Then, the charge carriers fully reach a state of localization and there is no transport.}
%     \label{Fig:Disorder Toymodel1}
% \end{figure}
Separating the effects of correlation and disorder is a highly non-trivial task. Generally, correlations seem to have a larger impact on the thermopower $S$ than disorder in \cobaltate. With the aim of understanding, we first set $U=0$ eV and gradually increase the disorder potential $\deps=0\rightarrow 1\,$eV. In the next section the opposite case, namely no disorder and increasing correlation, will be investigated.\\
The probability for a charge carrier to be scattered from one site \CoNa$\leftrightarrow$\CoVac\, to the other is obviously proportional to the doping $\xna$. On the other hand, the strength of the scattering event is determined by the disorder potential $\deps$. In fact, $\deps$ can be interpreted as an ``electron affinity'', since a larger $\deps$ will lead to more electrons occupying the sites \CoNa\ by means of the electrostatic attraction of the sodium ions. \\
%This effect is visualized in Fig.~\ref{Fig:Disorder Toymodel1}, where we show the qualitative change of the non-interacting density of states $N_\text{Na}$, $N_\text{Vac}$ of \CoNa\, and \CoVac, respectively. Generally, the centers of $N_\text{Na}$ and $N_\text{Vac}$ are shifted with respect to each other by $\OO(\deps)$, as the disorder potential $\deps$ is increased. At some point \nvac$\rightarrow0.5$, which will be important in the correlation analysis below. Additionally, the van-Hove-like peak of the less correlated site \CoNa\, crosses the Fermi level and thus leads to an increasing spectral weight for thermally activated holes. For very large disorder $\deps$ the system becomes insulating due to localization. \\
In Sec.~\ref{Sec:Num Self Energy and Spectrum}, the importance of the self energy $\Sigma(i\omega)$ on the imaginary axis has been discussed. With no correlation $U=0$, we can observe the pure scattering effects of the disorder. In Fig.~\ref{Fig:selfenergyandapproximativeformula1 T=290}, the imaginary part of the self energy $\Im \Sigma(i\omega)$ is shown. The value $\lim_{\omega\rightarrow 0}\Im\Sigma(i\omega)$ gradually increases as $\deps=0 \rightarrow 1\,$eV, which can be understood as follows:
%In the following, we fix the temperature $T=1160$ K, the doping $\xna$=0.7 and the disorder potential $\deps=0.55$ eV, and increase $U=0\rightarrow 5$ eV. The corresponding slopes of $\Im\Sigma(i\omega)$ can be found in Figure~\ref{Fig:selfenergy_spectrum_U1 T=1160}. The slopes of $\Im\Sigma(i\omega)$ change drastically with $U$ for small frequencies $\omega$. For small $U$, the charge carriers can be described as bare particle which are scattered at the disorder potential. As $U\rightarrow 5$ eV, the particles become dressed to form quasiparticles with a lifetime $\tau_L$. These quasiparticles can themselves be scattered by disorder.\\
%For small $\omega\rightarrow 0$, $\Im\Sigma(i\omega)$ converges to similar values, which are almost entirely determined by the disorder potential $\deps$. This can be seen as follows: Set $U=0$ for the moment and we aim to find $\lim_{\omega\rightarrow 0}\Im\Sigma(i\omega)$.
\begin{figure}[tp]
  \centering
 \includegraphics[height=7 cm]{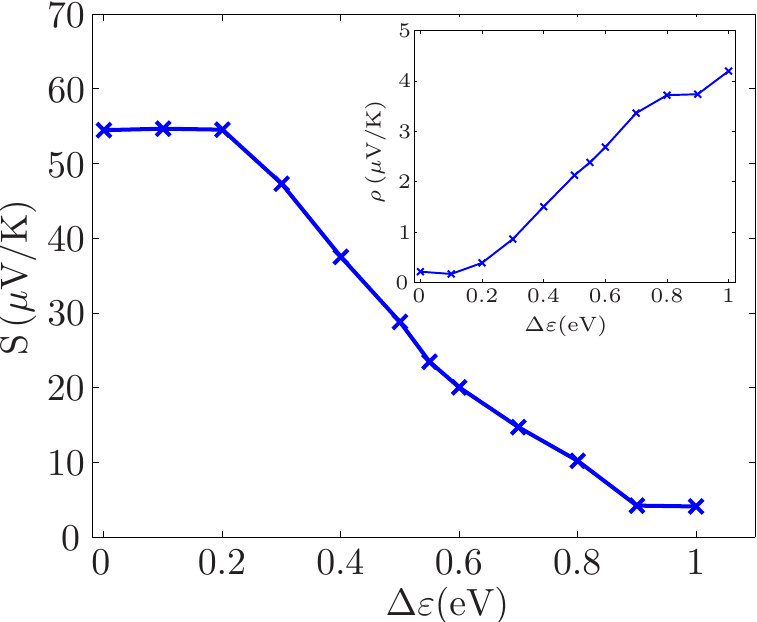}
    \caption{(color online) The thermopower $S$ as a function of the disorder potential $\deps$ without correlation ($U=0$) for $T=290$~K. The inset shows the corresponding values for the resistivity $\rho$. The magnitude of $S\sim K_1/K_0$ becomes smaller with larger $\deps$, which can be traced back to a decrease of the current-heat correlation function $K_1$ dominating the increasing $\rho=1/K_0$.}
    \label{Fig:SoverdepsU0 T=290}
\end{figure}
With the Greens function $G:=G(0)$ and the Weiss field $\Weiss^{-1}:=\Weiss^{-1}(0)$, we make a CPA~ansatz)
\begin{align}\label{Eq:CPA in results1}
 G = \frac{\xna}{\Weiss^{-1}+\deps}+\frac{1-\xna}{\Weiss^{-1}}.
\end{align}
This means that the total~(local) Greens function consists of contributions corresponding to the sub lattice \CoNa\, and \CoVac, where the former is energetically lowered by the disorder potential~$\deps$. As in the atomic limit, the task is to extract the self energy $\Sigma=\Sigma(0)$ by expressing $G$ of Eq.~\eqref{Eq:CPA in results1} with the corresponding Dyson equation
\begin{align}\label{Eq:CPA in results2}
 G = \frac{1}{\Weiss^{-1}-\Sigma}.
\end{align}
After some algebra, we find
\begin{align}\label{Eq:CPA in results3}
 %\Sigma = \frac{\xna\, \deps}{1-(1-\xna)\, \deps\, \Weiss}=\xna\, \deps \frac{(1-\xna)\, \deps\, \Re \Weiss+i(1-\xna)\deps\, \Im\Weiss}{[1-(1-\xna)\, \deps\, \Re \Weiss]^2+[(1-\xna)\,\deps\,\Im\Weiss]^2}
 \Sigma = \xna\, \deps \frac{(1-\xna)\, \deps\, \Re \Weiss+i(1-\xna)\deps\, \Im\Weiss}{[1-(1-\xna)\, \deps\, \Re \Weiss]^2+[(1-\xna)\,\deps\,\Im\Weiss]^2}
\end{align}
To obtain an approximation of the Weiss field $\Weiss^{-1}$, we use the Fourier transform of the non-interacting local Greens function $G_0(\omega=0)$
\begin{align}\label{Eq:CPA in results4}
\Weiss \approx =\PP\int d\varepsilon \frac{N(\varepsilon)}{-\varepsilon}+i\pi N(0),
\end{align}
where we used the Sokhatsky-Weierstrass theorem and $N(\varepsilon)$ denotes the non-interacting DOS. Note that we used the advanced version of the theorem, which means we will restrict to the part $\omega_m<0$ of the self energy $\Sigma(i\omega_m)$. Now, we assume a constant and symmetric DOS $N(\varepsilon)\approx N(0) \Theta(\varepsilon-\varepsilon_0)\Theta(\varepsilon_0-\varepsilon)$ with $N(0)\approx 0.8$\,eV$^{-1}$ from Fig.~\ref{Fig:DOS1}.
To that end, the principle value term drops in Eq.~\eqref{Eq:CPA in results4} and together with Eq.~\eqref{Eq:CPA in results3} we arrive at
\begin{align}\label{Eq:CPA in results5}
   \lim_{\omega\rightarrow 0}\Im\Sigma(i\omega)\approx \frac{\pi\, \xna(1-\xna)\, \deps^2\,  N(0)}{1+[\pi\,(1-\xna)\,\deps\, N(0)]^2}.
\end{align}
The values of $\lim_{\omega\rightarrow 0}\Im\Sigma(i\omega)$ as a function of $\deps$ are plotted as an inset in Fig.~\ref{Fig:selfenergyandapproximativeformula1 T=290}, and compared to the self-consistently determined DMFT+CPA results. The DMFT+CPA data indicate quadratic dependence with respect to $\deps$, which is properly approximated by our simple considerations.\\
The approximate formula \eqref{Eq:CPA in results5} indicates a direct connection between the disorder $\{\xna,\deps\}$ and the self energy at low frequency $\lim_{\omega\rightarrow 0}\Im\Sigma(i\omega)$. In fact, the value $\Im\Sigma(0)$ can be interpreted as an inverse time $\tau_\text{dis}^{-1}$, the average time which an electron spends at a site \CoNa\, or \CoVac, before scattering to the respective other site.\\
The increased scattering rate due to disorder is affecting both $K_0$ and $K_1$. Numerical results indicate that $K_1$ is stronger suppressed than the conductivity $K_0$ which corresponds to an overall drastic decrease of the thermopower $S$ which is visualized in Fig.~\ref{Fig:SoverdepsU0 T=290}. 
\newpage
\subsection{Effects of electronic correlation}\label{Sec:Correlation Effects}
\begin{figure}[tp]
  \centering
  \includegraphics[height=6.5 cm]{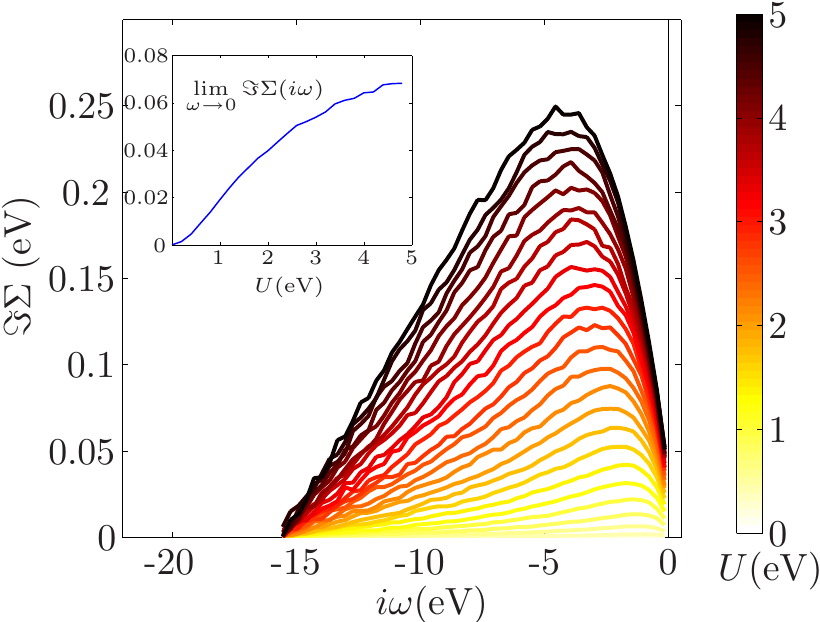}
    \caption{(color online) The imaginary part of the self energy $\Im\Sigma(i\omega)$ on the Matsubara axis for different values of the correlation $U=0\rightarrow 5\,$eV, and $T=290$~K~(no disorder). The actual values of $U$ are according to the colorbar on the right-hand side. (inset)~The values of $\lim_{\omega\rightarrow 0}\Im\Sigma(i\omega)$ as a function of $U$.}
    \label{Fig:KotliarVollhardt Mott-trans}
\end{figure}
After considering disorder without correlation in the previous section, we now focus on the effects of electron-electron correlation and assume that there is no disorder. At half filling $n/2=1/2$, it is well known that for increasing correlation $U$, a Mott metal-insulator transition can be observed~(see e.g. Ref.~\onlinecite{DMFT2}). \\%This transition, which is entirely driven by electronic effects, is introduced in the following.\\
Away from half filling $n/2<1/2$, an electron occupying a certain site simply hops to a neighboring empty site without the necessity of a double occupancy~(which would cost the energy~$U$). Thus, we  additionally obtain standard conduction channels. An equivalent picture is also applicable for holes for $n/2>1/2$ as in our model material \cobaltate. %Then, $U$ is already stored multiple times in the system, and has to be accounted for already at the crystallization of the material.\\
Without disorder, we have a mean filling of $n/2=0.85$, which is rather far away from $n/2=1/2$. For $U=3.5$\,eV, a renormalization of the non-interacting energy band of $Z=0.77$ is obtained. Consequently, without disorder, electronic correlations are only intermediately strong. However, it pays to investigate the effect of correlation on the self energy to compare with the case of combined disorder and correlation. In Fig.~\ref{Fig:KotliarVollhardt Mott-trans}, we plot the imaginary part of the self energy $\Sigma(i\omega)$ for different values of the correlation $U$. Several features can be observed: First, the maximum $\max\{\Im\Sigma(i\omega)\}\sim U$. Second, the limit $\lim_{\omega\rightarrow 0}\Im\Sigma(i\omega)\neq 0$ and the value depends on $U$. This behavior can be explained by means of Landau's Fermi Liquid Theory, which predicts a dependence $\lim_{\omega\rightarrow 0}\Im\Sigma(i\omega)\propto \alpha(U)T^2$. A visualization of the values $\lim_{\omega\rightarrow 0}\Im\Sigma(i\omega)$ as a function of $U$ can be found in the inset in Fig.~\ref{Fig:KotliarVollhardt Mott-trans}.\\
\begin{figure}[tp]
  \centering
 \includegraphics[height=6.5 cm]{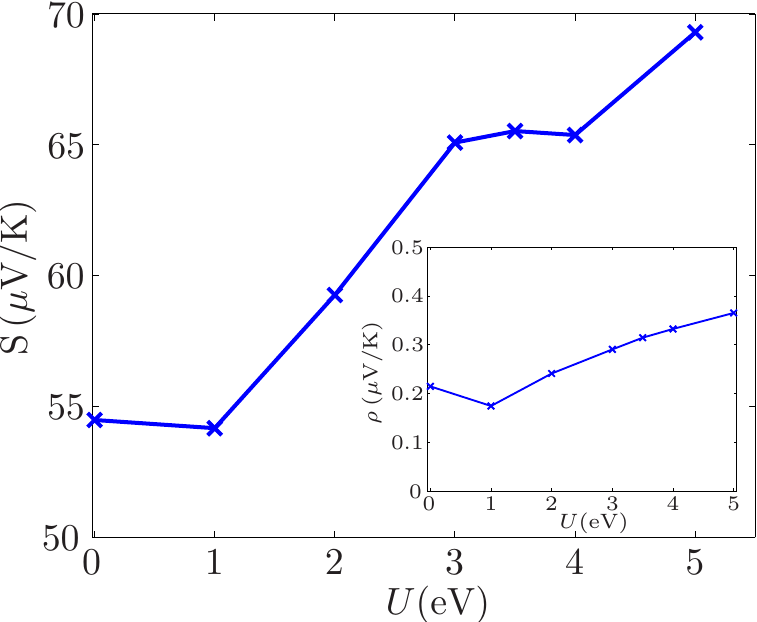}
    \caption{(color online) The thermopower $S$ as a function of the correlation $U$ without disorder ($\deps=0$) for $T=290$ K. The inset shows the corresponding values for the resistivity $\rho$. }
    \label{Fig:SoverUdeps0 T=290}
\end{figure}
In Fig.~\ref{Fig:SoverUdeps0 T=290}, we show the thermopower and the resistivity as a function of $U$ with $\deps=0$. Larger $U$ appears to increase $\rho$ which reflects on the thermopower $S$. However, the change of $S$ is much smaller than in the case of non-zero disorder $\deps>0$ below.\\
Correlation effects are small without the simultaneous consideration of disorder. When we include disorder, one of the two sites, namely \CoVac, will be driven closer to half filling. Then, the formation of a quasiparticle peak can be observed, which eventually leads to a higher thermopower.

\subsection{Effects of combined disorder and correlation}\label{Sec:Combined Effects}
\begin{figure*}[p]%Akw_N1.7_B40_U3.5_deps0.55.pdf
  \centering
\includegraphics[height=20.0 cm]{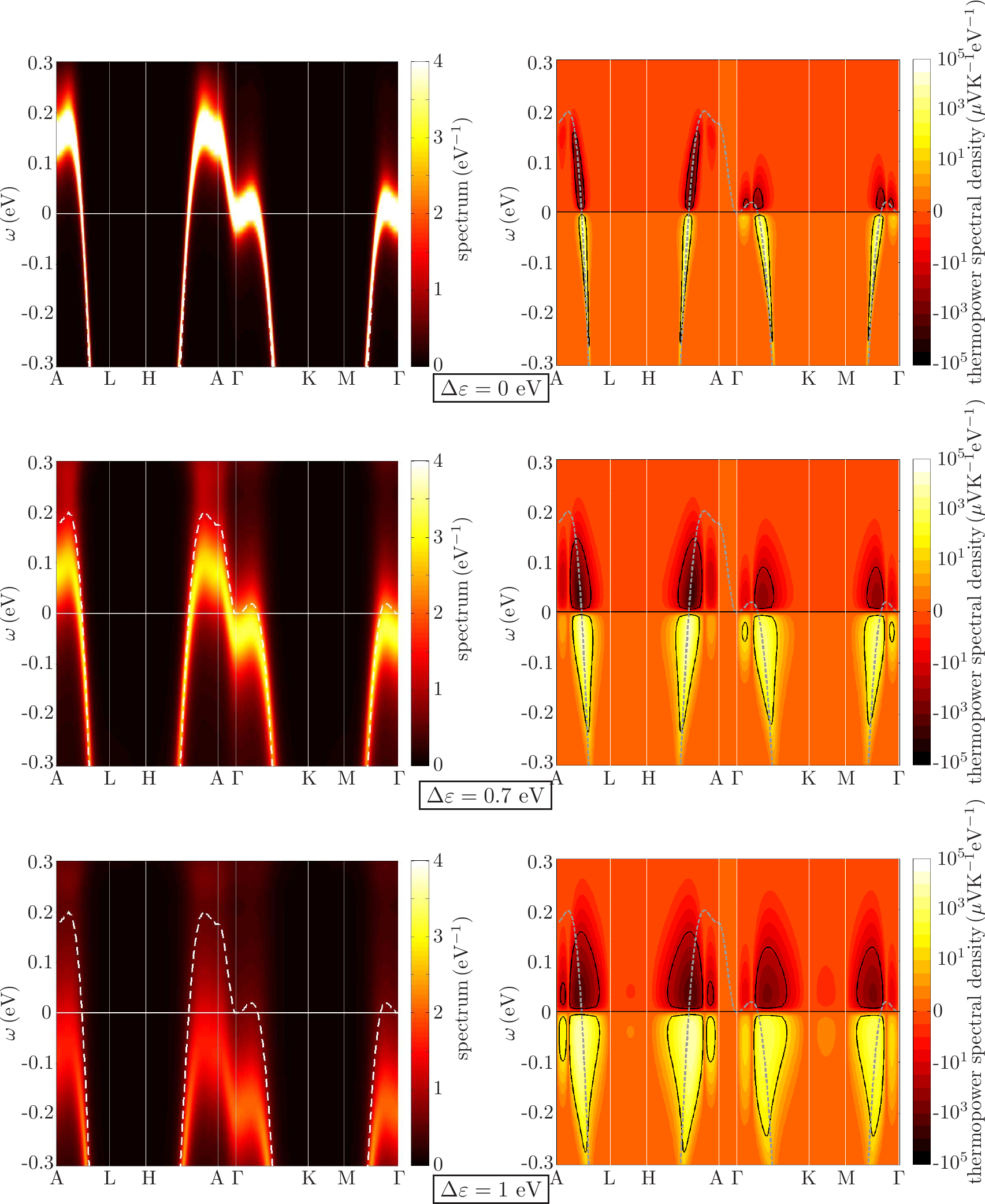}
    \caption{(color online) k-resolved spectra $A(k,\omega)$~(left) and corresponding thermopower spectral densities  \hspace*{0.05cm} $\KK_\text{tot}$~(right) for various values for the disorder potential $\deps=\{0\text{eV},0.7\text{eV},1\text{eV}\}$ at $T=290$ K. The color code is given by the colorbar and the tight-binding fit is visualized as dashed line. The electronic contribution $\omega>0$ to the thermopower is increased for $\deps=0.7$\,eV, but the hole contribution $\omega<0$ is enhanced even more. For large values of $\deps=1$\,eV spectral weight is diminishing above \emph{and} below the Fermi edge.}
    \label{Fig:Spectralcontributions_disorder}
\end{figure*}

\begin{figure*}[p]%Akw_N1.7_B40_U3.5_deps0.55.pdf
  \centering
 \includegraphics[height=20.0 cm]{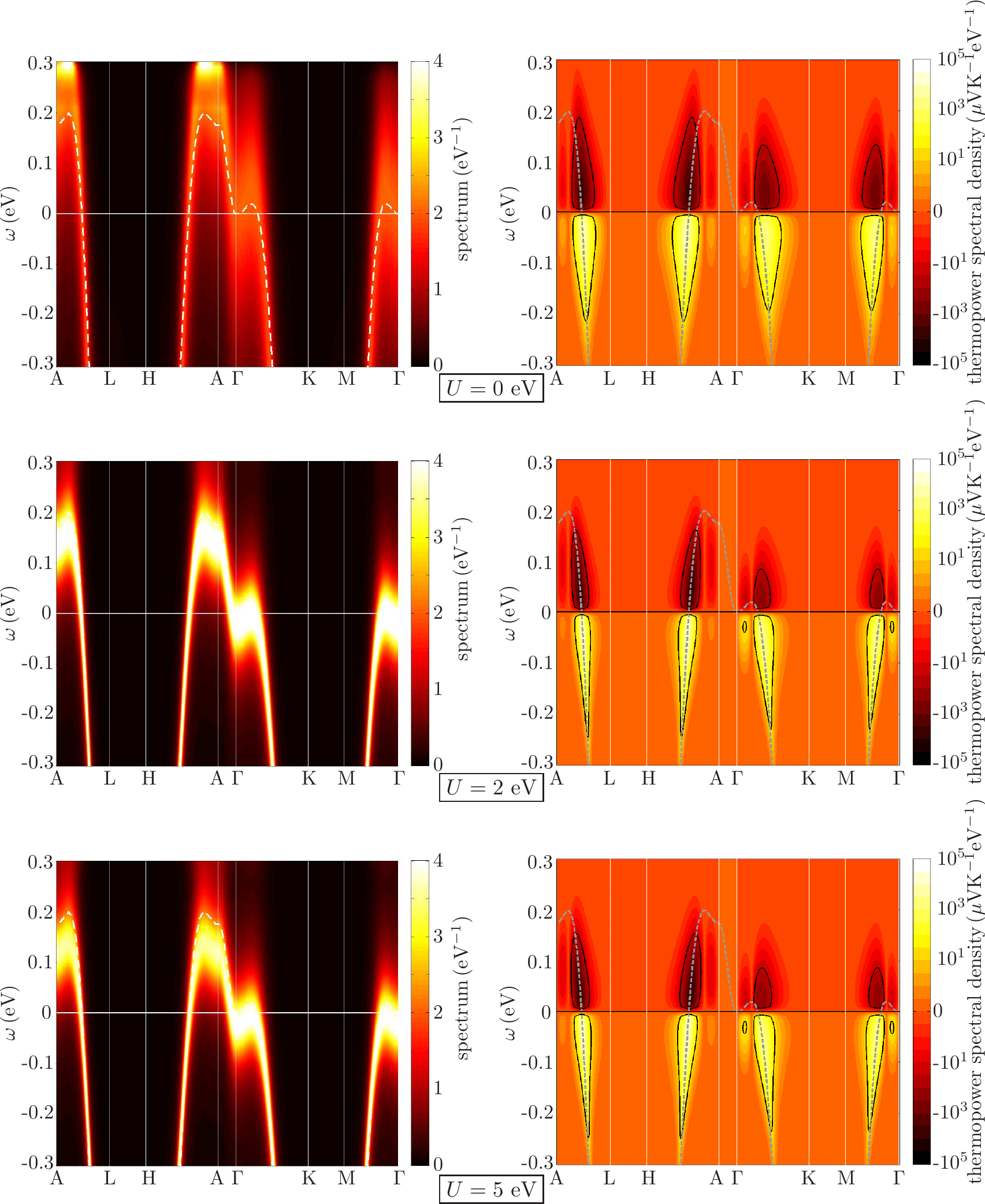}
    \caption{(color online) k-resolved spectra $A(k,\omega)$~(left) and corresponding thermopower spectral densities \hspace*{0.05cm} $\KK_\text{tot}$~(right) for various values for the correlation $U=\{0\ \text{eV},2\ \text{eV},5\ \text{eV}\}$ at $T=290$ K. The color code is given by the colorbar and the tight-binding fit is shown as a dashed line. Spectral weight is shifted from above, directly below the Fermi edge $\omega=0$ as $U=0\rightarrow2\,$eV, which increases the hole contributions to $S$ while decreasing the electron contribution. This effect is even stronger for $U=5$\ eV.}
    \label{Fig:Spectralcontributions_correlation}
\end{figure*}

The pure effects of disorder and correlation on self energy and spectra were discussed in the previous two sections. Whether the observed characteristics of disorder and correlation remain qualitatively unchanged in the presence of the other parameter, i.e. correlation or disorder, respectively, is yet to be investigated. We therefore fix either $U=3.5$\ eV or $\deps=0.55$\ eV and vary the respective other parameter~($T=290$ K and $\xna=0.7$).\\
\begin{figure}[tp]
  \centering
  \includegraphics[height=6.5 cm]{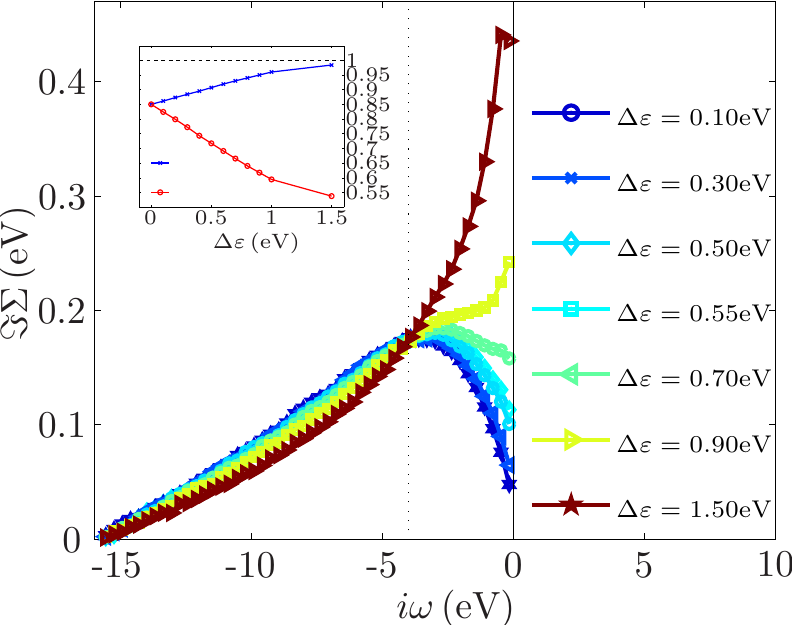}
    \caption{(color online) The imaginary part of the self energy $\Im\Sigma(i\omega)$ on the Matsubara axis for $T=290$~K, $U=3.5$\ eV and increasing values of the disorder potential $\deps$.(inset)~The filling per spin \nna,\nvac\, at the two sites \CoNa,\CoVac, respectively, as a function of the disorder potential $\deps$.}
    \label{Fig:selfenergy_combined T=290 disorder}
\end{figure}
In Fig.~\ref{Fig:selfenergy_combined T=290 disorder}, the imaginary part of the self energy $\Im\Sigma(i\omega)$ for different values of the disorder potential $\deps=0\rightarrow 1.5\,$eV and fixed $U=3.5$\ eV is shown. Two main energy regions can be identified: In the high energy region $\omega\lesssim-4$\ eV the behavior of the self energy is governed by the correlation $U$, which again can be explained by sum rules, cf. Ref.~\onlinecite{sumrule1}. On the other hand, between $-4$\ eV and $0$\ eV, the slope of $\Im\Sigma(i\omega)$ changes decisively with increasing $\deps$. However, the latter part of the slope may again be explained by the arguments of Sec.~\ref{Sec:Disorder Effects}, since the values of $\lim_{\omega\rightarrow 0}\Im\Sigma(i\omega)$ again depends approximately quadratically on the disorder~$\deps$ and agrees for small $\deps$. For $\deps=1.5$\ eV\,$\gtrsim t$, the system is already insulating, which is emphasized by an almost divergent $\Im\Sigma$ in the vicinity of $\omega=0$.\\
\begin{figure}[tp]
  \centering
  \includegraphics[height=11.0 cm]{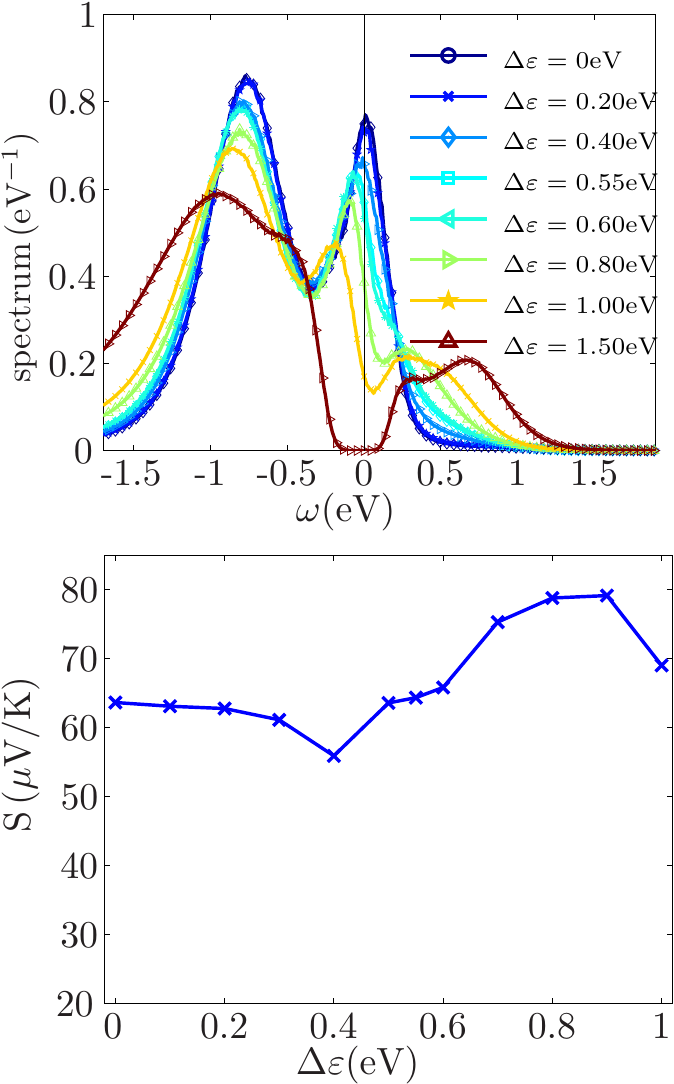}
    \caption{(color online) Top: k-integrated spectra for $U=3.5$\ eV, $T=290$ K and increasing disorder potential $\deps$. Spectral weight at the Fermi edge $\omega=0$ diminishes with increasing $\deps$, but at the same time the asymmetry is increased. Bottom:~The thermopower $S$ as a function of the disorder potential $\deps$ for a temperature $T=290$ K. Starting from $S=64\,\mu V/K$ for $\deps=0$\ eV the thermopower reaches a maximum $S=79\,\mu V/K$ for $\deps=0.9$\ eV.  For larger $\deps>0.9$ eV, a continuous decrease of $S$ can be observed as the system becomes insulating.}
    \label{Fig:spectra_combined disorder T=290}
\end{figure}
%\begin{figure}[t]
%  \centering
%  \begin{picture}(400,150) \thicklines \put(0,0){
%\psfrag{xna}{\small$\xna$}\psfrag{xva}{\small$\xvac$}
%  \put(100,0){\epsfig{file=NaCoO2_noverdeps1.pdf,height=5.5cm,clip=,angle=0}}
%  }
%\end{picture}
    %\caption{The filling $\xna$ and $\xvac$ of the two sublattice \CoNa and \CoVac, respectively, as a function of the disorder potential $\deps$. %As $\deps$ increases the filling reach their extremal values \nna=1 and \nvac=0.5, respectively. The system is then insulating.}
%    \label{Fig:fillingvsdeps1}
%\end{figure}
Another interesting numerical result is the filling of the a$_{1g}$ orbital on the two sub lattices \CoNa\, and \CoVac\, as a function of $\deps$, as visualized as an inset in Fig.~\ref{Fig:selfenergy_combined T=290 disorder}. The filling \nna\, rises in the analyzed $\deps$-interval basically linearly to 1 as the disorder potential $\deps$ increases, whereas \nvac\, decreases. For very large disorder $\deps=1.5$\ eV the fillings approach their maximum and minimal values \nna=$1$ and \nvac=$0.5$, respectively.\\
After the self energy, we investigate the spectrum for increasing disorder potential $\deps$. The k-integrated spectra $A(\omega)$ can be found in Fig.~\ref{Fig:spectra_combined disorder T=290}~(top). For $\deps=0\rightarrow 1\,$eV the spectral weight at the Fermi edge decreases, but the asymmetry of the spectral distribution with respect to $\omega=0$ increases. A further increase $\deps\gtrsim1$\ eV leads to the formation of a band gap and a alloy band splitting created by disorder.\\
Though less spectral weight implies decreasing thermopower contributions for both electrons and holes~(see $\KK_\text{tot}$ in (\ref{Eq:KubosformulainResults Kernelstot}) and Fig.\ref{Fig:Summary1}), asymmetry in the spectrum with respect to the Fermi edge $\omega=0$ enhances $S$, and the thermopower thus non-trivially depends on the disorder. In Fig.~\ref{Fig:spectra_combined disorder T=290}~(bottom), we show the numerical results for the thermopower $S$ as a function of the disorder potential $\deps$. The slope shows a flat maximum of the thermopower $S=80 \mu$V/K for $\deps\sim0.7$\ eV. For larger $\deps\gtrsim 0.9$\ eV, transport gets more and more suppressed as the spectral weight is shifted away from $\omega=0$ an alloy band gap forms. Again, the k-resolved spectra and contributions to the thermopower shown in Fig.~\ref{Fig:Spectralcontributions_disorder}, enhance the understanding of these effects: For no disorder $\deps=0$\ eV, we obtain a weak renormalization as discussed in Sec.~\ref{Sec:Correlation Effects}. With a disorder of  $\deps=0.7$\ eV all the charge carriers get scattered, which can be seen in a more diffuse k-resolved spectrum~(corresponding to a larger $\Im\Sigma(i\omega_m)$). However, as the corresponding contribution to the thermopower reveals, electrons become scattered stronger than holes. Therefore the net enhancement for the transport property $S$ in Fig.~\ref{Fig:spectra_combined disorder T=290}~(bottom), can be explained by an increase of spectral weight for the holes~$\omega<0$ compared to the purely disordered model, which is larger than the, also observed, increase of spectral weight for electrons~($\omega>0$). But for $\deps>0.8$\ eV the contribution of holes to the thermopower starts to diminish by a larger amount than can be accounted for by less negative contribution of the electrons. All charge carriers then are scattered at a high rate, the thermopower $S$ consequently drops as the system becomes insulating. \\
\begin{figure}[tp]
  \centering
 \includegraphics[height=6.5 cm]{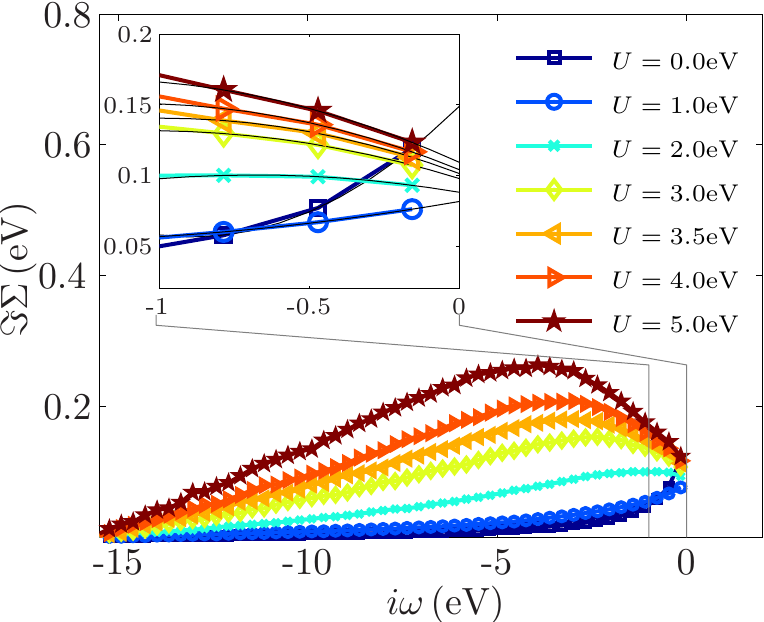}
    \caption{(color online) The imaginary part of the self energy $\Im\Sigma(i\omega)$ on the Matsubara axis for $T=290$ K, $\deps=0.55$ eV and increasing values of the correlation $U$. The black lines in the magnification are parabolic fits of the slopes. Note that the slopes converge to similar values for $\omega\rightarrow 0$, yet increase with larger $U$. For $U<\deps$, the value of $\lim_{\omega\rightarrow 0}\Im\Sigma(i\omega)$ appears to be governed by the disorder, while for larger $U$ the shape of $\Im\Sigma(i\omega)$ is mainly determined by correlation effects.}
    \label{Fig:selfenergy_combined T=290 correlation}
\end{figure} 
\begin{figure}[tp]
  \centering
 \includegraphics[height=12 cm]{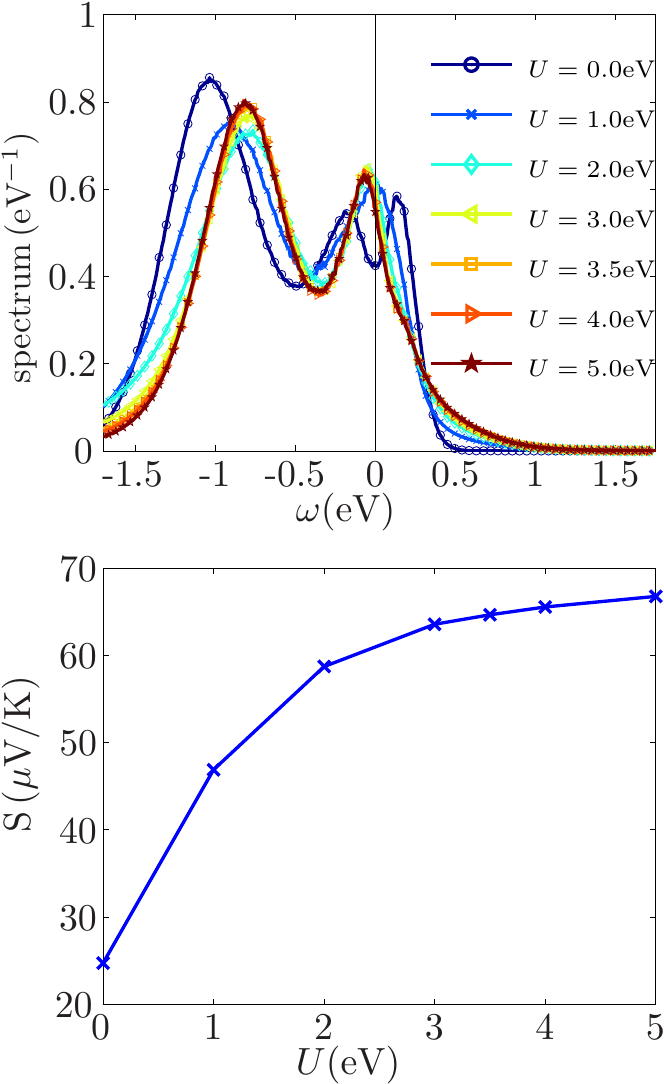}
    \caption{(color online) Top: k-integrated spectra for $\deps=0.55$ eV, $T=290$ K and increasing correlation $U$. Spectral weight is shifted from right above to right below the Fermi level $\omega=0$. Bottom:~The thermopower $S$ as a function of the correlation $U$ for a temperature $T=290$ K. The thermopower increases from $S\sim 25\mu\,V/K$ for $U=0$ eV up to a maximal value of $S=67\mu\,V/K$ for $U\sim 4$ eV, where the effect it saturated. }
    \label{Fig:spectra_combined correlation T=290}
\end{figure}
%\newpage\noindent
%An alternative explanation relying on the relative shift of spectral weight, can be found in Sec.~\ref{Sec:Disorder Effects}.\\
After the numerical results for increasing disorder, we now fix $\deps=0.55$\ eV and gradually increase the correlation $U=0\rightarrow 5\,$eV. In Fig.~\ref{Fig:selfenergy_combined T=290 correlation}, we plot the corresponding imaginary part $\Im\Sigma(i\omega)$. Surprisingly, the value $\lim_{\omega\rightarrow 0} \Im\Sigma(i\omega)$ first drops from $0.15$\ eV below $0.1$\ eV and then stabilizes around $0.1 $\ eV for increasing correlation $U$. This indicates a rather non-trivial impact of a changing $U$ on an already disorder system~(the situation is further complicated since the Luttinger theorem does not hold due to disorder, and non-zero temperature, and the van-Hove like peak around the Fermi edge appears to react very sensitive to changes of $U$).\\
The effect of increasing correlation on the k-integrated spectrum is shown in Fig.~\ref{Fig:spectra_combined correlation T=290}~(top). A shift of spectral weight from just above to just below the Fermi level can be observed. This indicates an increase of the thermopower for larger $U$, which is indeed found as it is shown in Fig.~\ref{Fig:spectra_combined correlation T=290}~(bottom), where we show $S$ function of $U$. The enhancement of $S=25\rightarrow 64 \mu$V/K is extreme as $U=0\rightarrow 3\,$eV. However, the growth of $S$ gets saturated for $U\gtrsim 5$\ eV. \\
To further improve our understanding, we again investigate the k-resolved spectra and corresponding contributions to the thermopower, cf. Fig.~\ref{Fig:Spectralcontributions_correlation}. Without correlation $U=0$, there is little spectral weight directly around the Fermi edge. However, in that energy interval a larger $U$ effectively packs spectral weight to larger extent for holes than for electrons. Thus, the drastic enhancement of $S$ can be understood. For larger $U\sim 5$ eV, the change at the Fermi edge is not decisive, resulting in the saturation of the thermopower $S$ with respect to $U$ shown in Fig.~\ref{Fig:spectra_combined correlation T=290}(bottom).
\begin{figure}[tp]
  \centering
   \includegraphics[height=6.5 cm]{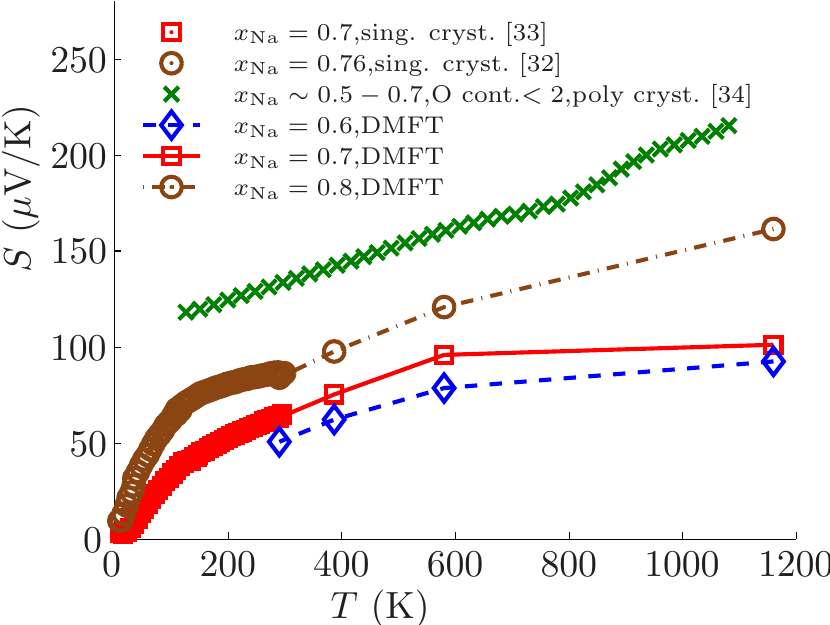}
    \caption{(color online) The thermopower $S$ over the temperature $T$ for $U=3.5$\ eV, $\deps=0.55$\ eV and various doping $\xna$ computed by DMFT compared to experiment. Transport in xy-direction is assumed, $\vv_{xy}^2=\vv_x^2+\vv_y^2$. The thermopower increases non-linearly with temperature $T$ as discussed in Sec.~\ref{Sec:Temperature Effects}. For increasing doping $\xna$, the thermopower increases for all temperatures~$T$.}
    \label{Fig:Seebeck1_x}
\end{figure}
\begin{figure}[t]
  \centering
  \includegraphics[height=12 cm]{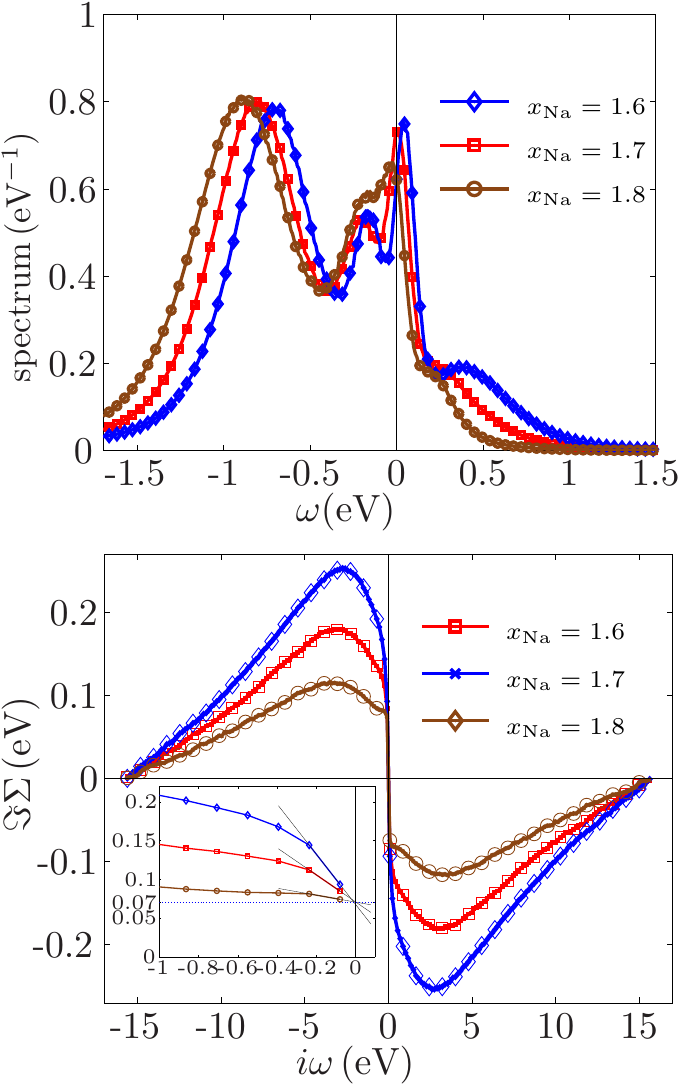}
    \caption{(color online) k-integrated spectra for $U=3.5$\ eV, $\deps=0.55$\ eV, $T=290$ K and various doping~$\xna$. For $\xna=0.6$ the Fermi liquid peak is the most pronounced as the filling of the site \CoVac\, is closest to $1/2$. In the vicinity of the Fermi edge, spectral weight is shifted from $\omega>0$ to $\omega<0$ as $\xna=0.6\rightarrow 0.8$.(bottom) The imaginary part of the self energy $\Im\Sigma(i\omega)$ on the Matsubara axis for $U=3.5$~eV, $\deps=0.55$\ eV, $T=290$ K and various doping $\xna=\{0.6,0.7,0.8\}$. (inset)~Magnification around $\omega=0$ and corresponding tangents~(black). All slopes converge to similar values which are independent of the disorder $\xna$, but dependent on the disorder potential $\deps$ and $U$.}
    \label{Fig:spectrum_doping1}
\end{figure}
\begin{figure*}[p]%Akw_N1.7_B40_U3.5_deps0.55.pdf
  \centering
\includegraphics[height=20.0 cm]{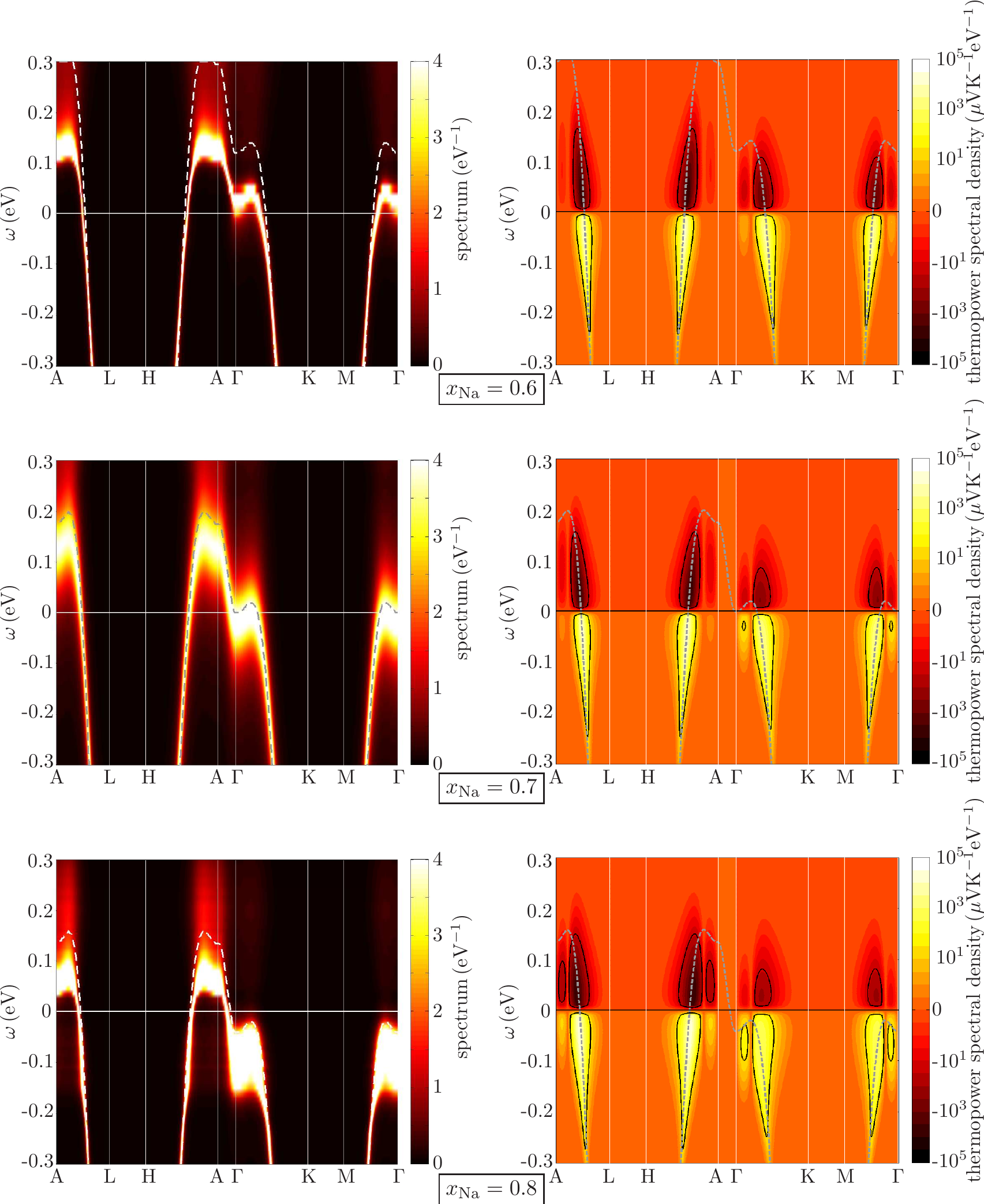}
    \caption{(color online) k-resolved spectra $A(k,\omega)$~(left) and corresponding thermopower spectral densities \hspace*{0.05cm}  $\KK_\text{tot}$~(right) for $U=3.5$\ eV, $\deps=0.55$\ eV and various values of the doping $\xna$. The color code is given by the colorbar and the tight-binding fit is visualized by a dashed line. The~(negative) electronic contribution to the thermopower is smaller as $\xna=0.6 \rightarrow 0.8$, thus the thermopower $S$ shows a net increase. For $\xna=0.8$ also the ``pudding molds'' between $A\rightarrow H$ and $A\rightarrow L$ becomes important.}
    \label{Fig:Spectralcontributions_doping}
\end{figure*}
%\clearpage

\subsection{Effects of the doping}\label{Sec:Thermopower for various dopings}
The considerations of the previous sections are mostly thought experiments to identify the important effects in the model compound \cobaltate. On the other hand, the typical external parameters one can control in experiment, are the temperature $T$ and the doping $\xna$. The gradual change or replacement of specific elements within a chemical compound has been used since ancient times. During the course of the 20th century, doping has become a major tool in technical applications and experimental materials science. \\
In semiconductor applications, one usually dopes by replacement: In addition to the main contents~(e.g. Si), a similar element is added which leaves the crystal structure in principle unchanged and takes the position of a fraction of the main chemical component(s). In the band structure, this procedure creates new bands which are energetically close to the ones of the desired charge carrier, and can therefore be much easier thermally activated than in the pure compound.\\
In contrast to replacement, \cobaltate\; is doped by changing the sodium content, cf. Refs.~\onlinecite{Yakabe98,Lee06,Kaurav09}. Since the single electron of the alkali Na becomes a valence electron in the cobaltate CoO$_2$ layers, this corresponds to shifting the Fermi energy $E_\text{F}$ within the a$_{1g}$ band of the non-interacting band structure~(Fig.~\ref{Fig:TightBinding1}). \\
As discussed in Sec.~\ref{Sec:LinearResponse} and in Ref.~\onlinecite{Kuroki07}, the ``pudding mold''-like slope of the a$_{1g}$ band is expected to be a very important feature for a high thermopower $S$. The closer the Fermi energy $E_{F}$ is to the flat top of the band, the larger the effect will be. Then, one charge carrier highly outmatches the other in terms of group velocity $\nabla \varepsilon(k)$. We therefore expect the thermopower to increase for higher $\xna$. In fact, as can be seen in Fig.~\ref{Fig:Seebeck1_x}, the computations yield a larger thermopower $S$ as the doping $\xna$ increases from $0.6\rightarrow0.8$. Here, we plot the thermopower $S$ over temperature $T$ for various doping $\xna$ and compare to experiment~(see Ref.~\onlinecite{short} for a corresponding comparison of the resistivity). The experimental results are in good agreement with our data, even though a direct comparison is made difficult, by the~(in this case) unknown orientation of the crystal in the experiment. %For comparison with single-crystal data, the results shown in Fig.~\ref{Fig:Seebeck1_x} correspond to the use of the xy group velocity  $\vv_{xy}^2=\nabla\varepsilon(k)\vert_x^2+\nabla\varepsilon(k)\vert_y^2$ instead of $\vv^2=\abs{\nabla\varepsilon(k)}^2$ in Equation~\eqref{Eq:LinResp Aftermath Thermopower}. For comparison with polycrystaline samples, we also provide the result for a mean group velocity $\overline{v}=(\vv_x^2+\vv_y^2+\vv^2_z)/3$ in Figure~\ref{Fig:Seebeck1_mean}. Here, the thermopower is $\sim$30\% smaller than in the case of $\vv_{xy}$, since the decrease of $K_1$ is larger than the decrease of $K_0$ when $\vv_{xy}\rightarrow\overline{v}$, cf.~Equation~\eqref{Eq:Thermopower Main1}).\\
The result of larger thermopower $S$ with increasing $\xna$ can be qualitatively explained by band structure arguments. Still, there can be additional effects due to correlations and disorder. The k-integrated spectra in Fig.~\ref{Fig:spectrum_doping1} indicate that spectral weight is shifted from electrons $\omega>0$ to holes $\omega<0$ as $\xna=0.6\rightarrow 0.8$. This further enhances the ``pudding mold'' effect of the group velocity $\nabla\varepsilon$. The k-resolved spectra in Fig.~\ref{Fig:Spectralcontributions_doping} show another result: The ``pudding mold'' contribution of the k-path in the xy-plane $\Gamma\rightarrow K$ important for $\xna=0.6,0.7$ is joined by the $k_z\neq 0$ contribution $A\rightarrow H$ as the Fermi level moves down. That provides an explanation why the dampening of the linear increase of the thermopower $S$ becomes smaller as $\xna=0.7\rightarrow0.8$.
\newpage
\section{Conclusion}\label{Sec:Conclusion}
In Na$_x$CoO$_2$ two distinct effects enhance the hole current in comparison
to the electron one: a larger group velocity and a larger number of
holes or,
more precisely, a larger hole spectral weight~\cite{quasi},
also see Fig.~\ref{Fig:Summary1}. As both effects point in the same direction, we
obtain a large electron-hole imbalance. Consequently, a temperature gradient results in a higher hole
current, and, hence, a large positive thermopower S. The larger group velocity of the
holes can be understood already from a one-particle LDA point of view. It is
caused by the particular pudding-mold type of bandstructure~\cite{Kuroki07}.
In this paper we analyzed the second mechanism which is a genuine
correlation effect. The microscopic origin is a complex interplay of Coulomb
repulsion $U$ and a (disordered) potential $\Delta \epsilon$ due to the Na vacancies. 
We have hence taken $U$ and $\Delta \epsilon$ as free parameters (not
{\em ab initio} ones as in Ref.~\onlinecite{short}). The Coulomb repulsion
alone (without disorder) has only a rather minute impact. This is due to
the filling of 1.7 electrons per Co site. With 0.3 holes in an otherwise
filled Co a$_{1g}$ orbital, chances to have 2 holes on the same Co site
are very small. Hence the impact of the Coulomb repulsion, whose main
effect is to  suppress such configurations, is small. Without the vacancy
potential, electronic
correlations in  Na$_x$CoO$_2$ would therefore be very weak.
Without Coulomb repulsion on the other hand, the vacancy disorder
smears out the spectrum and particularly enhances the resistivity less than
the current--heat-current correlation function is diminished, so that the thermopower
actually decreases. The story becomes very different if Coulomb repulsion
and disorder potential are combined. Then, the disorder potential gives
rise to two inequivalent lattice sites. Since the number of holes
in the $a_{1g}$ exactly corresponds to the number of sites with
a higher potential for the electrons,  we end up with a situation
close to two electrons on the Co sites with adjacent  Na-ion
and one  electron on the Co site with adjacent vacancy. In this situation, electronic
correlations can be very strong for the half-filled Co sites with an adjacent vacancy.
Because of this, we get a strongly renormalized spectrum with a sharp
peak at the Fermi level, see Fig.~\ref{Fig:Spectrum2}. The other sites,
with adjacent Na-ions, on the other hand, have only a sizeable spectral
weight on the hole-side of the Fermi level. Taken together, we have
more (quasi-)holes  but nonetheless sharp peaks in the spectrum.
In this situation, the thermopower increases up to 80 $\mu V/K$ with interaction.
For a larger value of the disorder potential, the Coulomb interaction
can even (counter intuitively) change the behavior from an insulator
with a very small thermopower to a metal with a big one.\\
For tracing down the energy and momentum of the quasiparticles
responsible for positive and negative thermopower alike, we have
analyzed the various contributions to the thermopower separately.
In Fig.~\ref{Fig:Spectralcontributions1}, we show for which  $k$ and  $\omega$
we have a significant contribution from the group velocity,
the spectral weight, and  for which interval holes and electrons
can be thermally activated. We also introduced a thermoelectric spectral
weight which combines these terms and hence provides for a $k$- and 
$\omega$-resolved analysis of the thermopower. The biggest contributions
naturally stem from those $k$-vectors which have the largest group velocity,
albeit the quantitative contribution is changing in comparison to
the non-interacting case. Most interestingly, also hole contributions
appear which, without interaction, were not present at certain $k$-points.
Generally, electronic correlations change the energies and momenta
with big contributions to the thermopower quite substantially.\\
As  the vacancy (disorder) potential $\Delta \epsilon$ is most
crucial for the thermopower and the physical properties of
Na$_x$CoO$_2$ in general, it is worthwhile to have a closer inspection
of this parameter. Actually, in LDA calculations, $\Delta \epsilon$ is rather
small, i.e., $\sim 0.05\,$eV(Ref.~\onlinecite{Singh00}). This value is way too small
to provide for the aforementioned charge disproportionation and the
resulting strong electronic correlations on the Co sites with vacancy.
Marianetti and Kotliar \cite{Kotliar07,footnote2},  argued that LDA overestimates
screening so that the actually $\Delta \epsilon$ is much larger, i.e.,
$\sim 0.55\,$eV. This is also the value we have considered in our
paper as the physically relevant one. Besides the overestimation
of screening, we think that the electron-phonon coupling is an important source
for the effective increase of  $\Delta \epsilon$ by $\sim g^2/\omega_0$
($g$: electron
phonon coupling strength; $\omega_0$: phonon frequency).

We acknowledge the financial support from the FWF through GK W004~(PW), from the research unit FOR 1346~(AT),
from the FWF through the ``Lise-Meitner'' Grant No. M1136~(GS), and from the SFB ViCoM F41~(KH). Calculations have been done on the Vienna Scientific Cluster~(VSC).

%\Myacknowledgement

\onecolumngrid


\begin{thebibliography}{10}

%\bibitem{Mahan}
%G.\ D.\ Mahan,  Solid State Physics {\bf 51}, 81 (1997).
%G.\ D.\ Mahan {\em et al.}, Physics Today, March 1997, p.42.

\bibitem{MRS} T. M. Tritt, M.~ A.~ Subramanian, MRS Bulletin, {\bf 31}, 188 (2006).

\bibitem{solar} M. Xie, D.M. Gruen, to appear on J. Phys. Chem. B (2010).

\bibitem{cars} J. Fairbanks, ``Thermoelectric applications in vehicles status 2008'' in ECT 2008 -- On line proceedings [available at ect2008.icmpe.cnrs.fr].

\bibitem{PbTeDOS}
%Enhancement of Thermoelectric Efficiency in PbTe by Distortion of the Electronic Density of States
J. P. Heremans  {\em et al.}, Science {\bf 321}, 554 (2008).% V. Jovovic, E. S. Toberer, A. Saramat, K. Kurosaki, A. Charoenphakdee, S. Yamanaka, and G. J. Snyder,

\bibitem{Paschen} S. Paschen, Thermoelectric aspects of strongly
correlated electron systems, in CRC Handbook of Thermoelectrics, Ch. 15,
(ed. D. M. Rowe, CRC Press), Boca Raton, 2005.

%\bibitem{transportnanostructures}
%J. K. Freericks, \emph{Transport in multilayered nanostructures: the dynamical mean-field theory approach}, Imperial College Press, %London, 2006.
\bibitem{TEPAM1}
J. K. Freericks \emph{et al}, %V. Zlati\'c, and A. M. Shvaika
 Phys. Rev. B {\bf 75}, 035133 (2007).

\bibitem{hetero1}
A.I. Hochbaum {\em et al.}, Nature {\bf 451}, 163 (2008)

\bibitem{hetero2}
A.I. Boukai {\em et al.}, Nature {\bf 451}, 168 (2008)

\bibitem{short}
P. Wissgott, A. Toschi, H. Usui, K. Kuroki, and K. Held, Phys. Rev. B {\bf 82}, 201106(R) (2010)
%{\sl Enhancement of the NaxCoO2 thermopower due to electronic correlations},

\bibitem{Singh00} D.J. Singh, Phys. Rev. B {\bf 61},  13397 (2000).

\bibitem{Kuroki07}
K.\ Kuroki, R.\ Arita, J. Phys. Soc. Jpn. {\bf 76}, 083707 (2007). Also note a related double pudding mold in LiRh$_2$O$_4$, R. Arita {\em et al.} PRB {\bf 78}, 115121 (2008)

\bibitem{DMFT1} W.\ Metzner and D.\ Vollhardt, Phys. Rev. Lett. {\bf 62},  324  (1989).

\bibitem{DMFT2}
A.\ Georges and G. Kotliar, Phys. Rev. B, {\bf 45},  6479 (1992).

\bibitem{DMFT3}
A.\ Georges et al, Rev. Mod. Phys. {\bf  68}, 13 (1996).%, G.~Kotliar, W.~Krauth and M.~Rozenberg

\bibitem{LDADMFT0}
V.\ I.\ Anisimov, A.~I. Poteryaev, M.~A. Korotin, A.~O. Anokhin, and G.~Kotliar, J. Phys. Condens. Matter,%
{\bf 9}, 7359 (1997).%, A.~I. Poteryaev, M.~A. Korotin, A.~O. Anokhin, and G.~Kotliar, J. Phys. Condens. Matter

\bibitem{LDADMFT1}
 A.\ I.\ Lichtenstein, M.\ I.\ Katsnelson,
Phys. Rev. B {\bf 57}, 6884 (1998).

\bibitem{LDADMFT4}
K.\ Held, Adv. Phys. {\bf 56}, 829 (2007).

\bibitem{Held09}
K. Held \emph{et al.}, in ``Properties and applications of thermoelectric materials'', eds.V.Zlatic,A.Hewson, Springer, 2009.

\bibitem{Lechermann05}
Frank Lechermann, Silke Biermann, Antoine Georges, Progress of Theoretical Physics Supplement {\bf 160} 233 (2005)
%Interorbital charge transfers and Fermi-surface deformations in strongly correlated metals: models, BaVS$_3$ and Na$_{x}$CoO$_2$ 

\bibitem{Ishida05}
H. Ishida \emph{et al}, Phys. Rev. Lett. {\bf 94}, 196401 (2005).
%Effect of Dynamical Coulomb Correlations on the Fermi Surface of Na0.3CoO2

\bibitem{Liebsch08}
A. Liebsch, H. Ishida, Eur. Phys. J. B {\bf 61},405-411 (2008).

\bibitem{Lechermann09}
F. Lechermann, Phys. Rev. Lett. {\bf 102}, 046403 (2009)
%Correlation Effects on the Doped Triangular Lattice in View of the Physics of Sodium-Rich NaxCoO2

\bibitem{Kotliar07}
C.A.\ Marianetti, G.\ Kotliar, Phys. Rev. Lett. {\bf 98}, 176405 (2007).
%Na induced correlations in NaCoO2

\bibitem{quasi}
More precisely one should speak of quasi-electrons and quasi-holes.

%\bibitem{BiTePh}
%B. Poudel {\em et al.}, Science {\bf 320}, 634 (2008).% Q. Hao, Y. Ma, Y. Lan, A. Minnich, B. Yu, X. Yan, D. Wang, A. Muto, D. Vashaee, X. Chen, J. Liu, M. S. Dresselhaus, G. Chen, and Z. Ren

%\bibitem{satellite}
%F. Ritz, C. E. Peterson, IEEE, Aerospace Conference, 2004. Proceedings, {\bf 5}, 2957 (2004).

%\bibitem{Handbook} D.M. Rowe, ed.,``CRC Handbook of Thermoelectrics'', CRC Press, Boca Raton, Florida (1995) and references therein.

\bibitem{Hubbard63}
J. Hubbard, Proc. Roy. Soc.,London A 276 238 (1963)

% \bibitem{Georges96}
% Georges A. \emph{et al.}, Rev. Mod. Phys. \bf{68}, 13 (1996)
% Dynamical Mean-Field Theory of Strongly-Correlated Fermion Systems and the Limit of Inﬁnite Dimensions

\bibitem{Elliot73}
R.J. Elliott, J.A. Krumhansl, P.L. Leath, Rev. Mod. Phys. {\bf 46}, 3 (1973).
%The Theory and Properties of Randomly Disordered Crystals and Related Physical Systems

\bibitem{Vlaming91}
R. Vlaming, D. Vollhardt, Phys. Rev. B {\bf 45}, 4637 (1991).
%Controlled Mean-Field Theory for Disordered Electronic Systems: Single-particle Properties

\bibitem{Mahan90}
G.D. Mahan, {\em Many-Particle Physics}, 2nd edition, Plenum Press, New York and London (1990)

\bibitem{footnote3}
This intrinsic difference between conductance and thermopower is due to the fact that the electric field $E$ distinguishes between electrons and holes, whereas the temperature gradient $\nabla T$, the driving force for thermopower, creates diffusion in the same direction of both electrons and holes.

\bibitem{Durst09}
Durst A.C. and Lee P.A., Phys.Rev. B {\bf 62}, 1270 (2000)
% title = {{Impurity-Induced Quasiparticle Transport and Universal Limit Wiedemann-Franz Violation in d-Wave Superconductors}},

\bibitem{Terasaki97}
%I.\ Terasaki, Y.\ Sasago and K.\ Uchinokura,
I.\ Terasaki \emph{et al}
Phys. Rev. B {\bf 56}, R12685(1997).

\bibitem{Kaurav09}
%N. Kaurav, K.K. Wu, Y.K. Kuo, Phys. Rev. B {\bf 79},075105 (2009).
N. Kaurav \emph{et al}, Phys. Rev. B {\bf 79},075105 (2009).
%Seebeck coefficient of Na$_{x}$CoO$_2$: Measurements and a narrow-band model

\bibitem{Lee06}
M. Lee  {\em et al.}, Nature Materials {\bf 5}, 537 (2006).
%Large enhancement of the thermopower in Na$_x$CoO$_2$ at high doping

\bibitem{Yakabe98}
Yakabe H. \emph{et al.}, 17th Int.Conf.o.Thermoelectrics, 551 (1998)
% title = {{Thermoelectric properties of Na$_{x}$CoO$_{2-\delta}$ system; Focusing on Partially Substituting Effects}},

\bibitem{Schaak03}
Schaak R.E. \emph{et al.},nature {\bf 424}, 527-529 (2003)
% Superconductivity phase diagram of Na$_x$CoO$_2$·1.3H$_2$O}},

\bibitem{Suguira06}
Suguira K. \emph{et al.}, Inorg. Chem. {\bf 45}, 5  1894-1896 (2006)
% {{Epitaxial Film Growth and Superconducting Behavior of
% Sodium−Cobalt Oxyhydrate, Na$_x$CoO·yH$_2$O (x=0.3, y= 1.3)}},


\bibitem{Balsys97}
Balsys R.J.  and {Davis} R.L., Solid State Ionics {\bf 93}, 279-282 (1997)
% title = {{Refinement of the structure of Na0.74CoO2 using neutron powder diffraction}},

\bibitem{Huang04}
Huang Q. \emph{et al.}, Phys. Rev. B {\bf 70}, 134115 (2004)
% Structural transition in Na$_x$CoO$_2$ with x near 0.75 due to Na rearrangement}},


\bibitem{Zandbergen04}
Zandbergen H.W. \emph{et al.}, Phys. Rev. B {\bf 70}, 024101 (2004)
% title = {{Sodium ion ordering in \Cobaltate: Electron diffraction study}},

\bibitem{footnote1} Some experiments\cite{Balsys97}
indicate the migration of the Na-ions, and, at lower temperatures,
the Na-ions might also form a superstructure, see Ref.~\cite{Zandbergen04}.

\bibitem{footnote2}
The strength of the disorder potential $\deps$ is very sensitive to screening effects and thus depends on the localization of the $t_{2g}$ electrons~\cite{Kotliar07}. In the LDA calculation, where screening is overestimated, $\deps$ is much smaller than 0.55 eV. However, correlation effects may lead to an even larger $\deps$. The value $\deps=0.55$ from Ref.~\onlinecite{Kotliar07} is thus an intermediate choice between these two extremes. In a real \emph{ab initio} approach, one would have to determine the disorder potential $\deps$ self-consistently~(i.e. including correlations).

\bibitem{ulmke}
The Fourier transforms needed for the DMFT cycle have been performed using the Ulmke smoothing scheme which implies that Im($\Sigma$) shown here artificially vanishes at the Nyquist frequency, see M. Ulmke, V. Janis and D. Vollhardt, Phys. Rev. B {\bf51}, 10411 (1995)

%\bibitem{Arita08}
%R. Arita {\em et al.}, Phys. Rev. B {\bf 78}, 115121 (2008).%, K. Kuroki, K. Held, A. V. Lukoyanov, S. Skornyakov, and
% V.I. Anisimov,

%\bibitem{Okamoto}
%Y.\ Okamoto  {\em et al.}, Phys. Rev. Lett. {\bf 101}, 086404 (2008).%S. Niitaka, M. Uchida, T. Waki, M. Takigawa, Y. Nakatsu, A. Sekiyama, S. Suga, R. Arita, and H. Takagi

% A. Bentien {\em et al.}, Europhys. Lett. {\bf 80}, 17008 (2007).

%\bibitem{FeSb2}
%A. Bentien et al,
% Europhys. Lett. {\bf 80}, 39901 (2007).% S. Johnsen, G. K. H. Madsen, B. B. Iversen, and F. Steglich

%\bibitem{TEPAM2}
%V. Zlati\'c, R. Monnier, and J. K. Freericks,
%Phys. Rev. B {\bf 78}, 045113 (2008).

%\bibitem{DFT}
%P.~Hohenberg and W.~Kohn, \newblock Phys. Rev. {\bf 136} B864 (1964).

%\bibitem{Wilson} G.B. Wilson-Short, D. J. Singh,
%M. Fornari, and M. Suewattana, Phys. Rev. B
%{\bf 75}, 035121  (2007).

%\bibitem{LDA}
%For  reviews see, e.g., R.~O. Jones and O.~Gunnarsson, \newblock Rev. Mod. Phys. {\bf 61} 689 (1989);
%R.~M. Martin, \newblock {\em Electronic Structure: Basic Theory and Practical Methods\/}
%  (Cambridge University Press, 2004).

%\bibitem{Nekrasov00}
%I.~A. Nekrasov, K.~Held, N.~Bl\"umer, A.~I. Poteryaev, V.~I. Anisimov and
%  D.~Vollhardt, \newblock Eur. Phys. J. B {\bf 18} 55 (2000).

%\bibitem{LDADMFT2}
%K.\ Held et al, phys. stat. sol. (b) {\bf 243}, 2599 (2006).%I.~A. Nekrasov, G.~Keller, V.~Eyert, N.~Bl\"umer, A.~McMahan, R.~Scalettar, T.~Pruschke, V.~I. Anisimov and D.~Vollhardt

%\bibitem{LDADMFT3}
%G. Kotliar, S.~Y. Savrasov, K.~Haule, V.~S. Oudovenko, O.~Parcollet, and C.~A.
%  Marianetti, Rev. Mod. Phys. {\bf 78}, 865 (2006).

%\bibitem{Perroni07}
%C.A. Perroni, H. Ishida, A. Liebsch, Phys. Rev. B {\bf 75} 045125 (2007).
%Exact Diagonalization Dynamical Mean Field Theory for Multi-Band Materials: Effect of Coulomb correlations on the Fermi surface of Na_0.3CoO_2

 %\bibitem{Hasan04}%Fermi Surface and Quasiparticle Dynamics of Na0:7 CoO2 Investigated by Angle-Resolved Photoemission Spectroscopy
%M.Z.\ Hasan \emph{et al}, Phys. Rev. Lett. {\bf 92}, 246402 (2004).
%M. Z. Hasan,1,2,3,* Y.-D. Chuang,1,3 D. Qian,1 Y.W. Li,1 Y. Kong,1 A. Kuprin,3 A.V. Fedorov,3 R. Kimmerling,3
      %E. Rotenberg,3 K. Rossnagel,3 Z. Hussain,3 H. Koh,3 N. S. Rogado,2,4 M. L. Foo,2,4 and R. J. Cava 2,4

\bibitem{Skilling89}
Skilling J., Fundamental theoris of physics {\bf 36}, Dordrecht Kluwer, Cambridge England (1989)
% title = {Maximum entropy and Bayesian methods},

\bibitem{Avella07}
Jarrel M.\emph{et al.}, Lectures on the Physics of Strongly Correlated Systems XII, AIP Conference Proceedings {\bf 1014},
ed.: Avella A. and Mancini F., Melville New York (2007)

\bibitem{Yang04}
H.-B. Yang \emph{et al}, Phys. Rev. Lett. {\bf 92},246403 (2004).
%ARPES on Na0.6CoO2: Fermi Surface and Unusual Band Dispersion

%\bibitem{Yang05}
%H.-B. Yang et al, Phys. Rev. Lett. {\bf 95},146401 (2005).
%Fermi Surface Evolution and Luttinger Theorem in NaxCoO2: A Systematic Photoemission Study

%\bibitem{notehopping} We obtain $t$=180 meV as nearest-neighbor hopping. Since hopping parameters to neighbors of higher order appear to be important for transport~\cite{Kuroki} the fit also included second- to fourth-nearest intraplanar neighbors t$'$= -39 meV, t$''$= -27 meV, t$'''$= 0.4 meV, and interplanar parameters beginning with the hopping to the prismatic Co neighbor up to the fourth interplanar neighbor, t$_{z0}=-18$ meV, t$_{z1}=-5$ meV, t$_{z1}=-2$ meV, t$_{z3}=1$ meV, t$_{z4}=0.5$ meV, respectively.

%\bibitem{Hirsch}
% J.~E. Hirsch and R.~M. Fye, \newblock Phys. Rev. Lett. {\bf 56}, 2521 (1986).

%\bibitem{KohnSham}
%W.~Kohn and L.~J. Sham, \newblock Phys. Rev. {\bf 140}, A1133 (1965).




% \bibitem{jellium}
%D.~M. Ceperley and B.~J. Alder, \newblock Phys. Rev. Lett. {\bf 45}, 566 (1980).

% \bibitem{Rubtsov04a}
% A.~N. Rubtsov and A.~I. Lichtenstein, \newblock JETP Lett. {\bf 80}, 61 (2004).
%
% \bibitem{Rubtsov05a}
% A.~N. Rubtsov, V.~V. Savkin and A.~I. Lichtenstein, \newblock Phys. Rev. B {\bf 72}, 035122 (2005).
%
% \bibitem{Werner05a}
% P.~Werner, A.~Comanac, L.~{De Medici}, M.~Troyer and A.~J. Millis, \newblock Phys. Rev. Lett. {\bf 97}, 076405 (2006).
%
% \bibitem{Werner06a}
% P.~Werner and A.~J. Millis, \newblock Phys. Rev. B {\bf 74}, 155107 (2006).
%
% \bibitem{Sakai}
% S.~Sakai, R.~Arita, K.~Held and K.~Aoki, \newblock Phys. Rev. B {\bf 74}, 155102 (2006).
%
% \bibitem{LDAU}
% V.~I. Anisimov, J.~Zaanen and O.~K. Andersen, \newblock Phys. Rev. B {\bf 44}, 943 (1991).
%
% \bibitem{spinpolaron}
% G.~Sangiovanni, A.~Toschi, E.~Koch, K.~Held, M.~Capone, C.~Castellani,
%   O.~Gunnarsson, S.-K. Mo, J.~W. Allen, H.-D. Kim, A.~Sekiyama, A.~Yamasaki,
%   S.~Suga and P.~Metcalf, \newblock Phys. Rev. B {\bf 73}, 205121 (2006).
%
%
% \bibitem{clusterDMFT1}
% T.~Maier, M.~Jarrell, T.~Pruschke and M.~H. Hettler, \newblock Rev. Mod. Phys. {\bf 77} 1027 (2005).
%
% \bibitem{clusterDMFT2}
% G.~Kotliar, S.~Y. Savrasov, G.~P\'alsson and G.~Biroli, \newblock Phys. Rev. Lett. {\bf 87} 186401 (2001).
%
% \bibitem{clusterDMFT3}
% A.~I. Lichtenstein and M.~I. Katsnelson, \newblock Phys. Rev. B {\bf 62} 9283 (R) (2000).
%
%
%
% \bibitem{DGA1}
% A.~Toschi, A.~A. Katanin and K.~Held, Phys. Rev. B 75, 045118 (2007).
%
%
% \bibitem{DGA2}
% H.~Kusunose, J. Phys. Soc. Jpn. {\bf 75}, 054713  (2006).
%
% \bibitem{DGA3}
% C.~Slezak, M.~Jarrell, T.~Maier and J.~Deisz  (2006), \newblock {cond-mat/0603421}.
%
% \bibitem{DGA4}
% K. Held, A. A. Katanin, A. Toschi, arXiv:0807.1860.
%
%
% \bibitem{Pruschke}
% T.~Pruschke, D.~L. Cox and M.~Jarrell, \newblock Phys. Rev. B {\bf 47} 3553 (1993).
%
% \bibitem{MaxEnt}
% M.~Jarrell and J.~E. Gubernatis, \newblock Physics Reports {\bf 269} 133 (1996).
% \bibitem{Nekrasov}
% I.\ Nekrasov et al, Phys. Rev. B {\bf 73}, 155112 (2006).
% \bibitem{Pade}
% H. J. Vidberg and J. W. Serene, J. Low Temp. Phys. {\bf 29}, 179 (1977)


% \bibitem{Koshibae}
% W.\ Koshibae, K. Tsutsui, and S. Maekawa, Phys. Rev. B {\bf 62} 6869 (2000).
% \bibitem{hotdebate} I. Terasaki, JPSJ Online-News and Comments [Oct. 10, 2007].
% %\bibitem{Ishida} Y. Ishida et al., J. Phys. Soc. Jpn. {\bf 76},  103709 (2007{\it et al.}).
%
% \bibitem{LMTO} O.\ K.\ Andersen, Phys. Rev. B {\bf 12}, 3060 (1975);
% O.\ Gunnarsson, O.\ Jepsen, and O.\ K.\ Andersen,
% Phys. Rev. B {\bf 27}, 7144 (1983)
%
% \bibitem{Vanderbilt}
% N. Marzari and D. Vanderbilt, Phys. Rev. B {\bf 56}, 12847 (1997).
%
% \bibitem{Projection}
% V.\ I.\ Anisimov,
% D. E. Kondakov, A. V. Kozhevnikov,
% I. A. Nekrasov, Z. V. Pchelkina, J. W. Allen, S.-K. Mo, H.-D. Kim,
% P. Metcalf, S. Suga, A. Sekiyama, G. Keller, I. Leonov, X. Ren,
% and D. Vollhardt
% Phys. Rev. B {\bf 71}, 125119 (2005).
% %\bibitem{Paul} I. Paul and G. Kotliar, Phys. Rev. B {\bf 67}, 115131 (2003).
% \bibitem{Pchelkina}
% Z.\ Pchelkina et al, Phys. Rev. B {\bf 75}, 035122 (2007).
%\bibitem{Okamoto2}
%Y. Okamoto, private communication.
%++++++++++++PW add

\bibitem{sumrule1}
Y. Vilk and A.-M. Tremblay,  J. Phys. I (France), 7, 1309 (1997).


%\bibitem{note1}
%We obtain $t=(180,-39,-27,0.4)$ meV as nearest- to fourth-nearest neighbor hopping in plane;
%$t_{z}=(-18,-5,-2,1,0.5)$ meV are the
% interplanar parameters from the first (prismatic) Co neighbor up to the fourth.

%\bibitem{Oudovenko}
%V.\ S.\ Oudovenko {\it et al.}, Phys. Rev. B {\bf 73}, 035120 (2006).%G. P\'alsson, K. Haule, G. Kotliar, and
%S. Y. Savrasov



%\bibitem{Wang04}
%N.L. Wang  \emph{et al}, Phys. Rev. Lett. {\bf 93},237007 (2004)

%************
%\bibitem{Sales04}
%B.C. Sales  \emph{et al}, Phys. Rev. B. {\bf 70},174419 (2004)
%************

%\bibitem{Tomczak10}
%J.M. Tomczak \emph{et al}, Phys. Rev. B 82, 085104 (2010) 
% Thermopower of correlated semiconductors: Application to FeAs2 and FeSb2

%\bibitem{Fuchs20}
%Sebastian Fuchs \emph{et al}, http://arxiv.org/abs/1012.5950 (2010)
% Spectral properties of the three-dimensional Hubbard model

%\bibitem{Foo04}
%M.L. Foo et al, Phys. Rev. Lett. {\bf 92},247001 (2004)
%Charge-ordering, commensurability and metallicity in the phase diagram of layered Na(x)CoO(2)

%\bibitem{Schulze08}
%T.F. Schulze et al, Phys. Rev. B {\bf 78},205101 (2008)
%Spin fluctuations, magnetic long-range order, and Fermi surface gapping in  NaxCoO2


%+++++++++++PW end add

  \end{thebibliography}
\end{document}